\documentclass[12pt]{article}
\usepackage{makeidx}
\usepackage{amssymb}
\usepackage{amsfonts}
\usepackage{amsmath}
\usepackage{graphicx}
\usepackage{setspace}
\usepackage{authblk}
\usepackage{units}
\usepackage{appendix}
\usepackage[left=2cm,top=2.3cm,right=2cm,bottom=2.1cm]{geometry}
\usepackage{xcolor}
\usepackage[hyperfootnotes=false]{hyperref}
\hypersetup{colorlinks=true,citecolor=cyan,urlcolor=Red}

\begin{document}

\title{{\textbf{Magnetic response from constant backgrounds to Coulomb
sources}}}
\author[1,2]{T. C. Adorno\thanks{adorno@hbu.edu.cn, tg.adorno@gmail.com}}
\author[2,3,4]{D. M. Gitman\thanks{gitman@if.usp.br}}
\author[2,3]{A. E. Shabad\thanks{shabad@lpi.ru}}
\affil[1]{\textit{Department of Physics, College of Physical Sciences and Technology, Hebei University, Wusidong Road 180, 071002, Baoding, China;}}
\affil[2]{\textit{Department of Physics, Tomsk State University, Lenin Prospekt 36, 634050, Tomsk, Russia;}}
\affil[3]{\textit{P. N. Lebedev Physical Institute, 53 Leninskiy prospekt,
119991, Moscow, Russia;}}
\affil[4]{\textit{Instituto de F\'{\i}sica, Universidade de S\~{a}o Paulo, Caixa Postal 66318, CEP 05508-090, S\~{a}o Paulo, S.P., Brazil;}}
\maketitle

\onehalfspacing

\begin{abstract}
Magnetically uncharged, magnetic linear response of the vacuum filled with
arbitrarily combined constant electric and magnetic fields to an imposed
static electric charge is found within general nonlinear electrodynamics.
When the electric charge is point-like and external fields are parallel, the
response found may be interpreted as a field of two point-like magnetic
charges of opposite polarity in one point. Coefficients characterizing the
magnetic response and induced currents are specialized to Quantum
Electrodynamics, where the nonlinearity is taken as that determined by the
Heisenberg-Euler effective Lagrangian.
\end{abstract}

\section{Introduction}

It is well understood that the vacuum filled with strong background field
is, in Quantum Electrodynamics (QED), equivalent to a linear or nonlinear
medium \cite{Erber,EulKoc35,Heisenberg}. The simplest (one-loop) Feynman
diagrams responsible for description of such a media are shown in Fig. \ref%
{Fig0.1}.

\begin{figure}[th!]
\begin{center}
\includegraphics[scale=0.4]{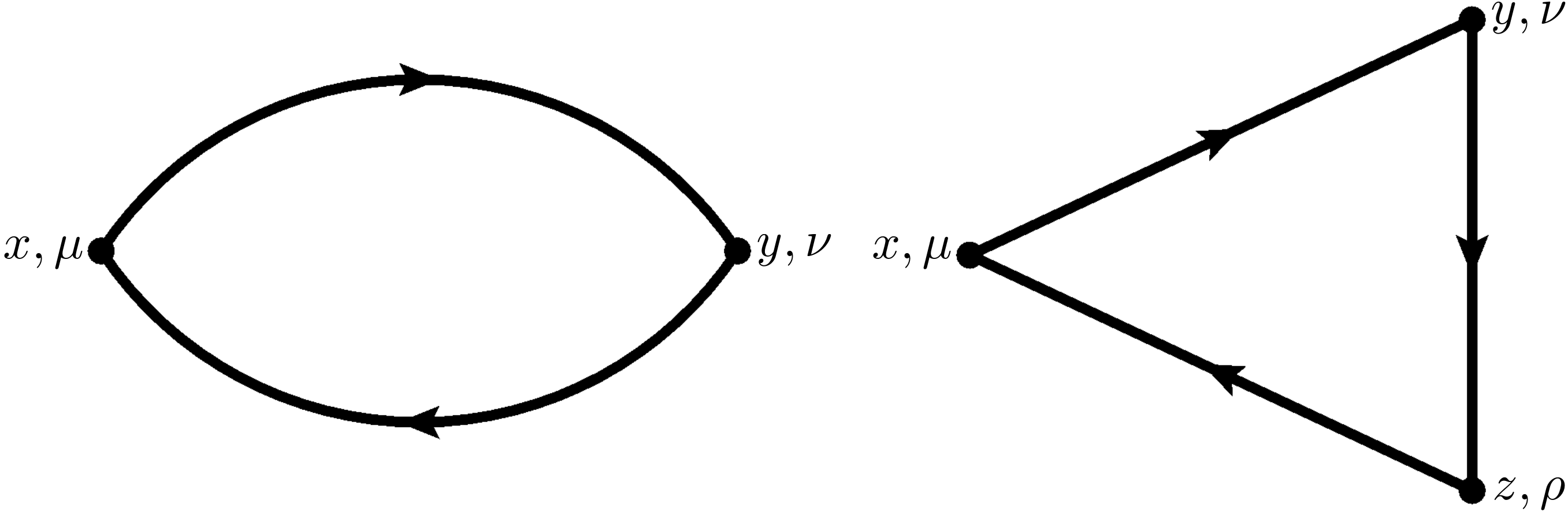}
\end{center}
\caption{The second- and third-rank polarization tensors.}
\label{Fig0.1}
\end{figure}

The bold lines there stand for the electron propagators in the external
field. These are known exactly for the constant background, for the
background plane electromagnetic field, also for special combinations of the
latter two. In the first case the equivalent medium is space- and
time-homogeneous, otherwise it is not, and then the energy and momentum
exchange between the field and the background occurs. Using the exact
solutions to the Dirac equation in external field makes these diagrams
belonging to the so-called Furry picture. The first diagram corresponds to
linearization of the field equations above the background. It represents the
(second-rank) polarization tensor $\Pi _{\mu \nu }(x,y),$ which contains in
itself the linear polarizational properties of the equivalent medium,
usually referred to as dielectric permeability and magnetic permittivity. It
is responsible for the screening of charges and currents and transformations
of their shapes due to the strong background, and for small-amplitude
electromagnetic wave propagations in the background, especially for
polarization of the eigen-modes and (different) modifications of the mass
shell in each mode (the birefringence making a goal for\ observation \cite%
{birefringence} with the recent evidence for it obtained from the\ neutron
star RX J1856.5-3764 \cite{birefringence2}) by deviating the dispersion
curves from the standard shape $k_{0}^{2}=\mathbf{k}^{2}$ known\ in the
empty vacuum (the one with null background). The second diagram in Fig. \ref%
{Fig0.1}, the third-rank polarization tensor $\Pi _{\mu \nu \rho }(x,y,z)$,
takes into account the quadratic response of the background. When taken on
the photon mass shell, it is responsible for the photon splitting and
merging in an external field \cite{Adler,merging}. Beyond the mass shell, it
also describes the response of the medium to small perturbations with the
quadratic accuracy relative to these perturbations. Analogously, the
fourth-rank polarization tensor includes the cubic response,
photon-by-photon scattering (the first experimental detection of this
fundamental process, which is the source of nonlinearity of QED, was
recently reported in \cite{light}), photon splitting into three \cite%
{merging}, and so on.

\begin{figure}[th!]
\begin{center}
\includegraphics[scale=0.4]{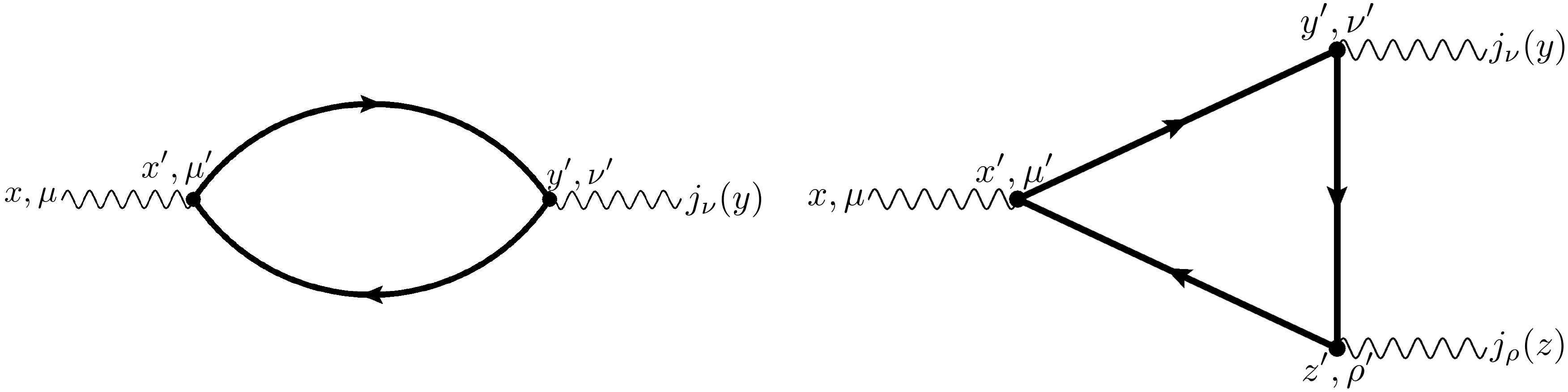}
\end{center}
\caption{The linear and quadratic responses to applied current $j_{\protect%
\mu }\left( x\right) $.}
\label{Fig0.2}
\end{figure}

The polarization tensors may be defined as variational derivatives of the
effective action $\Gamma $ known as \cite{weinberg} the generating
functional of the one-particle-irreducible vertex functions\footnote{%
Greek indices span the 4-dimensional Minkowski space-time, e. g., $\mu
=(0,i),\,i=1,2,3$, $\eta _{\mu \nu }=\mathrm{diag}(+1,-1,-1,-1),$ and
boldface letters denote three-dimensional Euclidean vectors (e. g., $\mathbf{%
A}(x)=\left( A^{i}(x)\,,\,\,i=1,2,3\right) $. The four-rank and three-rank
Levi-Civita tensors are normalized as $\varepsilon ^{0123}=+1$ and $%
\varepsilon _{123}=1$, respectively.},%
\begin{equation}
\Pi ^{\alpha \rho }\left( x,y\right) =\left. \frac{\delta ^{2}\Gamma }{%
\delta A_{\alpha }\left( x\right) \delta A_{\rho }\left( y\right) }%
\right\vert _{A=\overline{A}}\,,\text{\ \ }\Pi ^{\alpha \rho \beta }\left(
x,y,z\right) =\left. \frac{\delta ^{3}\Gamma }{\delta A_{\alpha }\left(
x\right) \delta A_{\rho }\left( y\right) \delta A_{\beta }\left( z\right) }%
\right\vert _{A=\overline{A}}\,,  \notag
\end{equation}%
with respect to the potentials $A_{\alpha }\left( x\right) $, taken at their
background values $A_{\alpha }\left( x\right) =\overline{A}_{\alpha }\left(
x\right) $. This is equivalent to the Feynman diagram representation. Each
differentiation adds an extra photon vertex to the diagram. The polarization
tensors of all ranks participate in the nonlinear Maxwell equations%
\begin{eqnarray}
&&\left[ \eta _{\rho \alpha }\square -\partial _{\rho }\partial _{\alpha }%
\right] a^{\rho }\left( x\right) +\int d^{4}y\Pi _{\alpha \rho }\left(
x,y\right) a^{\rho }\left( y\right)   \notag \\
&&+\frac{1}{2}\int d^{4}yd^{4}z\Pi _{\alpha \rho \beta }\left( x,y,z\right)
a^{\rho }\left( y\right) a^{\beta }\left( z\right) +O\left( a^{3}\right)
=j_{\alpha }\left( x\right) \,,  \label{equation}
\end{eqnarray}%
where $a^{\alpha }\left( x\right) $ is the potential over the background, $%
a_{\alpha }\left( x\right) =A_{\alpha }\left( x\right) -\overline{A}_{\alpha
}\left( x\right) $, i. e., the response to the applied source (perturbation) 
$j_{\rho }\left( x\right) $. Bearing in mind that each vertex in the
diagrams carries a small quantity, the electron charge $e$, we may approach
this equation perturbatively. Then, at the classical level, the solution to (%
\ref{equation}) is 
\begin{equation}
a_{(0)}^{\alpha }\left( x\right) =\int D_{(0)}^{\alpha \alpha ^{\prime
}}(x-x^{\prime })j_{\alpha ^{\prime }}\left( x^{\prime }\right)
d^{4}x^{\prime }\,,  \label{0}
\end{equation}%
where $D_{(0)}^{\nu \rho }(x-x^{\prime })$ is the free photon propagator.
The first correction to it (within the linearity of the equation) is%
\begin{equation}
a_{(1)}^{\alpha }\left( x\right) =\int D_{(0)}^{\alpha \alpha ^{\prime
}}(x-x^{\prime })\Pi _{\alpha ^{\prime }\rho ^{\prime }}\left( x^{\prime
},y^{\prime }\right) D_{(0)}^{\rho ^{\prime }\rho }(y^{\prime }-y)j_{\rho
}\left( y\right) d^{4}x^{\prime }d^{4}y^{\prime }d^{4}y\,.  \label{1}
\end{equation}%
The correction of the second power of the perturbation $j_{\rho }\left(
x\right) $ is 
\begin{equation}
a_{(2)}^{\alpha }\left( x\right) =\frac{1}{2}\int D_{(0)}^{\alpha \alpha
^{\prime }}(x-x^{\prime })\Pi _{\alpha ^{\prime }\rho ^{\prime }\beta
^{\prime }}\left( x^{\prime },y^{\prime },z^{\prime }\right) D_{(0)}^{\rho
^{\prime }\rho }(y^{\prime },y)j_{\rho }\left( y\right) D_{(0)}^{\beta
^{\prime }\beta }(z^{\prime },z)j_{\beta }\left( z\right) d^{4}x^{\prime
}d^{4}y^{\prime }d^{4}z^{\prime }d^{4}yd^{4}z\,.  \label{2}
\end{equation}%
The terms (\ref{1}) and (\ref{2}) are shown graphically in Fig. \ref{Fig0.2}%
, where the wiggly line stands for the free photon propagator $D_{(0)}^{\nu
\rho }(x-x^{\prime })$.

An essential simplification of the calculations is achieved if one confines
oneself to the approximation of the local effective action, where the
functional $\Gamma $ does not depend on the space-time derivatives of the
fields, for instance, where the Euler-Heisenberg expression for it is taken
in one-loop \cite{Heisenberg} approximation in QED\ (also two-loop \cite%
{ritus}, \cite{Karb} and even three-loop \cite{schubert}\ results are being
considered). This approximation is good as long as the fields slowly varying
in time and space are dealt with. Contrary to (\ref{equation}), within this
approximation the field equations are differential (not integral) ones, and
they do not include higher derivatives, while (the Fourier transform of) the
polarization tensor of $n$-th rank behaves as the $n$-th power of the
momentum; in the language of optics this corresponds to disregard of the
spatial and frequency dispersion\footnote{%
Various local Lagrangians that, like the Heisenberg-Euler approximation, do
not contain space- and time-derivatives of fields, but are not associated
with QED, are widely used together with the Einstein's gravity, especially
when studying magnetized black holes. See e.g. \cite{kruglov} and pertinent
references therein.}. We thoroughly traced the derivation of the Maxwell
equations within the local approximation in \cite{ShaGit2012}, \cite%
{CosGitSha2013}, \cite{AdoGitSha2013}, \cite{AdoGitShaShi2016}. In \cite%
{CosGitSha2013} the self-interaction of magnetic and electric dipole
moments, which modifies their values calculated within any version of the
strong-interaction theory, was considered using the 4-th rank polarization
tensor with no background. In \cite{CosGitSha2013a} the self-interaction of
a point-like charge was studied with the same tools leading to the result
that its field-energy (beyond the perturbation approach) is finite, although
the field in the center of the charge is not. Therefore, the point charge,
with its field being a solution of a nonlinear equation, becomes a soliton
at rest or in motion \cite{Shishmarev}, \cite{Ependiev}. Interaction between
long-wave electromagnetic waves was considered taking into account,
effectively, the polarization tensors up to 6-th rank \cite{King}, \cite{Di
Piazza}. In \cite{AdoGitSha2013}, \cite{AdoGitSha2014} we showed that the
quadratic response of the vacuum with the background of a constant magnetic
field to an applied electric field of a point-like or extended
central-symmetric charge, goverened by the 3-rd rank polarization tensor and
corresponding to term (\ref{2}) is purely magnetic, i.e. we face here the
magneto-electric effect. Moreover, the magnetic response far from the charge
is the field of a magnetic dipole with its dipole moment quadratically
dependent upon the electric charge. The photon splitting on the basis of the
same diagram was studied in \cite{merging}. In \cite{AdoGitSha2016} and now
we take an arbitrary combination of constant electric $\mathbf{\bar{E}}$ and
magnetic $\mathbf{\bar{B}}$ fields as a background (see \cite{BatSha71}
beyond the local approximation), and we consider linear response to an
applied electric charge following the information contained in the 2-nd rank
polarization tensor (second term in Eq. (\ref{equation})). This response may
be both electric and magnetic. The electric response was studied in \cite%
{AdoGitSha2016}, resulting in description of the induced charge density and
modification of the Coulomb field far from the charge\footnote{%
A brief review of our previous works may be found in \cite{AdoGitShaShi2016}.%
}. In the present paper we study the linear magnetic response of the
constant background to an applied Coulomb source, complementary to our
previous study \cite{AdoGitSha2016}, where only\textrm{\textbf{\ }}the
linear electric response was found. Linearly induced currents and vector
potentials are discussed in detail. Contrary to \cite{AdoGitSha2015} and to
our forthcoming work, only magnetic response with vanishing total magnetic
charge is considered here. Correspondingly, the finally found magnetic field
looks like a combination of two opposite point-like magnetic charges
coexisting in one point. Applying the general results derived for any
nonlinear theory to the special case where the nonlinearity is provided by
QED at one-loop, we study relevant coefficients characterizing the results
in terms of the Heisenberg-Euler effective action \cite{Heisenberg}, in
which proper-time representations and asymptotic regimes are discussed in
detail.

The paper is organized as follows. In Sec. \ref{Sec3}, after presenting the
necessary Maxwell equations linearized near the background field and
indicating the structure of the applied electric field, we obtain
expressions for the current density induced in the \textquotedblleft
medium\textquotedblright\ inside and ouside of the applied extended charge.
In Subsecs. \ref{Ss3.1} and \ref{S4} we find the magnetic fields produced by
this current and the vector potential corresponding to the magnetic response
produced by a pointlike Coulomb source. All the results reported above are
written in terms of the derivatives of the local effective Lagrangian over
the field invariants taken at the background. Hence these may be used with
every model Lagrangian, irrespective of its origin and of its connection to
QED. On the contrary, in Sec. \ref{S5}, we specialize the results to the
one-loop Euler-Heisenberg Lagrangian of QED. Sec. \ref{Conc} is devoted to
the concluding remarks.

\section{Linearly induced currents and magnetic responses in constant
backgrounds\label{Sec3}}

Let there be a background electromagnetic field, with its field tensor $%
F_{\nu \mu }\left( x\right) $ equal to $\overline{F}_{\nu \mu }\left(
x\right) $, produced by the background current $\mathcal{J}_{\mu }$ via the
(second set of) Maxwell equations%
\begin{equation}
\partial ^{\nu }\overline{F}_{\nu \mu }\left( x\right) -\partial ^{\nu }%
\left[ \left. \frac{\delta \mathfrak{L}\left( \mathfrak{F},\mathfrak{G}%
\right) }{\delta \mathfrak{F}\left( x\right) }\right\vert _{F=\overline{F}}%
\overline{F}_{\nu \mu }\left( x\right) +\left. \frac{\delta \mathfrak{L}%
\left( \mathfrak{F},\mathfrak{G}\right) }{\delta \mathfrak{G}\left( x\right) 
}\right\vert _{F=\overline{F}}\widetilde{\overline{F}}_{\nu \mu }\left(
x\right) \right] =\mathcal{J}_{\mu }\left( x\right) \,,
\label{background equation}
\end{equation}%
within a nonlinear local electrodynamics with the Lagrangian%
\begin{equation}
L(x)=-\mathfrak{F}\left( x\right) +\mathfrak{L(}x\mathfrak{)\,},  \label{L}
\end{equation}%
where its nonlinear part $\mathfrak{L(}x\mathfrak{)}$ is taken as a function 
$\mathfrak{L(}x)=\mathfrak{L}\left( \mathfrak{F},\mathfrak{G}\right) $ of
two field invariants $\mathfrak{F}=\left( 1/4\right) F^{\mu \nu }F_{\mu \nu
}=\left( 1/2\right) \left( \mathbf{B}^{2}-\mathbf{E}^{2}\right) $, $%
\mathfrak{G}=\left( 1/4\right) \tilde{F}^{\mu \nu }F_{\mu \nu }=-\left( 
\mathbf{E}\cdot \mathbf{B}\right) $, $\tilde{F}_{\mu \nu }\left( x\right)
=\left( 1/2\right) \varepsilon _{\mu \nu \alpha \beta }F^{\alpha \beta
}\left( x\right) $, and may be thought of, for instance, as the effective
Lagrangian of Quantum Electrodynamics in the local (infra-red)
approximation, i. e. the one where the dependence on the space- and
time-derivatives of the fields is neglected\footnote{%
The special case with the Euler-Heisenberg Lagrangian accepted as the
one-loop approximation of such local effective action will be treated for
the purposes of the present article in Section \ref{S5} below.}. Moreover,
we shall be considering the constant background $\overline{F}_{\nu \mu
}\left( x\right) =\overline{F}_{\nu \mu }=\mathrm{const.}$ here. This field
does not require any current to be supported: it is seen that equation (\ref%
{background equation}) is satisfied by $\mathcal{J}_{\mu }\left( x\right) =0$
in this case.

Let the constant background be disturbed by a small introduced current $%
j_{\mu }\left( x\right) $. It causes the deviation $f_{\nu \mu }\left(
x\right) =F_{\nu \mu }\left( x\right) -\overline{F}_{\nu \mu }$ of the field
from the background. Expanding the Maxwell equations in powers of $f_{\nu
\mu }\left( x\right) $ we obtain in the first order the linear equation (see
Refs. \cite%
{ShaGit2012,CosGitSha2013,AdoGitSha2013,AdoGitShaShi2016,CosGitSha2013a,AdoGitSha2014, AdoGitSha2016,AdoGitSha2015}
for equations for higher orders, which are nonlinear as containing the
second and higher powers of $f_{\nu \mu }\left( x\right) $):%
\begin{eqnarray}
\partial ^{\nu }f_{\nu \mu }\left( x\right) &=&j_{\mu }^{\mathrm{lin}}\left(
x\right) +j_{\mu }\left( x\right) \,,  \notag \\
j_{\mu }^{\mathrm{lin}}\left( x\right) &=&\partial ^{\tau }\left[ \mathfrak{L%
}_{\mathfrak{F}}f_{\tau \mu }\left( x\right) +\frac{1}{2}\left( \mathfrak{L}%
_{\mathfrak{FF}}\overline{F}_{\alpha \beta }+\mathfrak{L}_{\mathfrak{FG}}%
\widetilde{\overline{F}}_{\alpha \beta }\right) \overline{F}_{\tau \mu
}f^{\alpha \beta }\left( x\right) \right]  \notag \\
&+&\partial ^{\tau }\left[ \mathfrak{L}_{\mathfrak{G}}\tilde{f}_{\tau \mu
}\left( x\right) +\frac{1}{2}\left( \mathfrak{L}_{\mathfrak{FG}}\overline{F}%
_{\alpha \beta }+\mathfrak{L}_{\mathfrak{GG}}\widetilde{\overline{F}}%
_{\alpha \beta }\right) \widetilde{\overline{F}}_{\tau \mu }f^{\alpha \beta
}\left( x\right) \right] \,,  \label{s1.51}
\end{eqnarray}%
where the subscripts by $\mathfrak{L}$ designate derivatives with respect to
the indicated field invariants taken at their background value, for instance 
$\left. \frac{\partial ^{2}\mathfrak{L}}{\partial \mathfrak{F}\partial 
\mathfrak{G}}\right\vert _{F=\overline{F}}=\mathfrak{L}_{\mathfrak{FG}}$. We
have introduced here the notation for the linearly induced current $j_{\mu
}^{\mathrm{lin}}\left( x\right) $ (nonlinearly induced currents were dealt
with in \cite{ShaGit2012,CosGitSha2013,AdoGitShaShi2016,CosGitSha2013a,
AdoGitSha2016,AdoGitSha2013,AdoGitSha2014,AdoGitSha2015}). To avoid possible
misunderstanding, we stress that nonlinearly induced currents are
responsible for selfinteraction of the deviation fields $f_{\nu \mu }\left(
x\right) $, whereas the nonlinearity of the theory given by the Lagrangian (%
\ref{L}) shows itself in the present framework as the interaction between
the electromagnetic field $f_{\nu \mu }\left( x\right) $ and the
electromagnetic background $\overline{F}_{\alpha \beta }$).

In what follows we solve Eq. (\ref{s1.51}) perturbatively with respect to
the above coefficients, whose connection with QED will be exploited in
Section \ref{S5}. To this end we represent the electromagnetic
field-strength tensor as%
\begin{equation}
f_{\nu \mu }\left( x\right) =f_{\nu \mu }^{\left( 0\right) }\left( x\right)
+f_{\nu \mu }^{\left( 1\right) }\left( x\right) +...,  \label{pert}
\end{equation}%
where $f_{\nu \mu }^{\left( 0\right) }\left( x\right) $ is a solution of the
classical field equation%
\begin{equation*}
\partial ^{\nu }f_{\nu \mu }^{\left( 0\right) }\left( x\right) =j_{\mu
}\left( x\right) \,,
\end{equation*}%
while the linear response $f_{\nu \mu }^{\left( 1\right) }\left( x\right) $
is subject to the equation determined by the induced current taken on $%
f_{\nu \mu }^{\left( 0\right) }\left( x\right) $%
\begin{eqnarray}
\partial ^{\nu }f_{\nu \mu }^{\left( 1\right) }\left( x\right) &=&\partial
^{\nu }\left[ \frac{1}{2}\left( \mathfrak{L}_{\mathfrak{FF}}\overline{F}%
_{\alpha \beta }+\mathfrak{L}_{\mathfrak{FG}}\widetilde{\overline{F}}%
_{\alpha \beta }\right) \overline{F}_{\nu \mu }f^{\left( 0\right) \alpha
\beta }\left( x\right) +\mathfrak{L}_{\mathfrak{F}}f_{\nu \mu }^{\left(
0\right) }\left( x\right) \right]  \notag \\
&+&\partial ^{\nu }\left[ \frac{1}{2}\left( \mathfrak{L}_{\mathfrak{FG}}%
\overline{F}_{\alpha \beta }+\mathfrak{L}_{\mathfrak{GG}}\widetilde{\mathcal{%
F}}_{\alpha \beta }\right) \widetilde{\overline{F}}_{\nu \mu }f^{\left(
0\right) \alpha \beta }\left( x\right) +\mathfrak{L}_{\mathfrak{G}}\tilde{f}%
_{\nu \mu }^{\left( 0\right) }\left( x\right) \right] \,.  \label{17}
\end{eqnarray}%
For the perturbation of the background we take the current corresponding to
a static charge $q$ homogeneously distributed over a sphere\footnote{%
Hereafter $\theta \left( z\right) $ denotes the Heaviside step function
defined as $\theta \left( z\right) =1$ if $z\geq 0$ and zero otherwise.}
with the radius $R$%
\begin{eqnarray}
&&j_{\mu }\left( x\right) =\delta _{\mu 0}\rho ^{\left( 0\right) }\left(
r\right) \,,\ \ r=|\mathbf{r|\,,}  \notag \\
&&\rho ^{\left( 0\right) }\left( r\right) =\frac{3q}{4\pi R^{3}}\theta
\left( R-r\right) \,,\ \ R=\mathrm{const.}\,,  \label{const charge}
\end{eqnarray}%
This charge density corresponds to a regularization of the pointlike static
charge%
\begin{equation}
\rho ^{\left( 0\right) }\left( \mathbf{r}\right) =q\delta ^{3}\left( \mathbf{%
r}\right) \,,\ \ \delta ^{3}\left( \mathbf{r}\right) =\delta \left(
x^{1}\right) \delta \left( x^{2}\right) \delta \left( x^{3}\right) \,,
\label{pointlike}
\end{equation}%
placed at origin $\mathbf{r}=0$. It is a source of the regularized Coulomb
field $f_{0i}^{\left( 0\right) }\left( \mathbf{r}\right) =E^{\left( 0\right)
i}\left( \mathbf{r}\right) $ and null magnetic field $B^{\left( 0\right)
i}\left( \mathbf{r}\right) =-\left( 1/2\right) \varepsilon _{ijk}f^{\left(
0\right) jk}\left( \mathbf{r}\right) =0$%
\begin{eqnarray}
&&\partial ^{\nu }f_{\nu \mu }^{\left( 0\right) }\left( x\right) =j_{\mu
}\left( x\right) \,,\ \ \mathbf{E}^{(0)}\left( \mathbf{r}\right) =\mathbf{E}%
_{\mathrm{in}}^{\left( 0\right) }\left( \mathbf{r}\right) \theta \left(
R-r\right) +\mathbf{E}_{\mathrm{out}}^{\left( 0\right) }\left( \mathbf{r}%
\right) \theta \left( r-R\right) \,,  \notag \\
&&\mathbf{E}_{\mathrm{in}}^{\left( 0\right) }\left( \mathbf{r}\right) =\frac{%
q\mathbf{r}}{4\pi R^{3}}\,,\ \ \mathbf{E}_{\mathrm{out}}^{\left( 0\right)
}\left( \mathbf{r}\right) =\frac{q\mathbf{r}}{4\pi r^{3}}\,,\ \ \mathbf{B}%
^{\left( 0\right) }\left( \mathbf{r}\right) =0\,.  \label{in2}
\end{eqnarray}%
Throughout the text, the indexes \textquotedblleft in\textquotedblright\ and
\textquotedblleft out\textquotedblright\ classify electromagnetic quantities
at points inside $(r<R)$ and outside $(r\geq R)$ of the spherical charge
distribution, respectively.

In our previous work \cite{AdoGitSha2016},\ we studied the electric response 
$E^{\left( 1\right) k}(\mathbf{r)}\neq 0$,\newline
$B^{\left( 1\right) k}(\mathbf{r)}=-\left( 1/2\right) \varepsilon
_{ijk}f^{\left( 1\right) jk}\left( \mathbf{r}\right) =0$ to equation (\ref%
{17}) giving a correction to the Coulomb law (\ref{in2}); now we shall
consider the magnetic solution. Substituting the zero-order solutions $%
\mathbf{B}^{\left( 0\right) }\left( \mathbf{r}\right) =0$ and $\mathbf{E}%
^{(0)}\left( \mathbf{r}\right) $ (\ref{in2}) in Eq. (\ref{17}), one finds
that the first-order linear magnetic response $\mathbf{B}^{\left( 1\right) }(%
\mathbf{r)}$ to the purely electric perturbation (\ref{const charge}) is the
solution of the differential equation%
\begin{equation}
\boldsymbol{\nabla }\times \left[ \mathbf{B}^{\left( 1\right) }\left( 
\mathbf{r}\right) -\mathbf{\mathfrak{H}}^{(0)}\left( \mathbf{r}\right) %
\right] =0\,,  \label{nnew7b}
\end{equation}%
where $\mathbf{\mathfrak{H}}^{(0)}\left( \mathbf{r}\right) $ is the
expression within the brackets in (\ref{s1.51}), taken in the zeroth order 
\footnote{%
It should be noted that the relativistic covariance is deprived by selecting
the reference frame in which the static charge is at rest.}:%
\begin{eqnarray}
\mathbf{\mathfrak{H}}^{(0)}\left( \mathbf{r}\right) &=&-\mathfrak{L}_{%
\mathfrak{G}}\mathbf{E}^{(0)}\left( \mathbf{r}\right) -\left[ \mathfrak{L}_{%
\mathfrak{FF}}\mathbf{\overline{E}}\cdot \mathbf{E}^{(0)}\left( \mathbf{r}%
\right) +\mathfrak{L}_{\mathfrak{FG}}\mathbf{\overline{B}}\cdot \mathbf{E}%
^{(0)}\left( \mathbf{r}\right) \right] \mathbf{\overline{B}}  \notag \\
&+&\left[ \mathfrak{L}_{\mathfrak{FG}}\mathbf{\overline{E}}\cdot \mathbf{E}%
^{(0)}\left( \mathbf{r}\right) +\mathfrak{L}_{\mathfrak{GG}}\mathbf{%
\overline{B}}\cdot \mathbf{E}^{(0)}\left( \mathbf{r}\right) \right] \mathbf{%
\overline{E}}  \notag \\
&=&\mathbf{\mathfrak{H}}_{\mathrm{in}}^{\left( 0\right) }\left( \mathbf{r}%
\right) \theta (R-r)+\mathbf{\mathfrak{H}}_{\mathrm{out}}^{\left( 0\right)
}\left( \mathbf{r}\right) \theta (r-R)\,,  \label{nnew13b}
\end{eqnarray}%
in which%
\begin{eqnarray}
\mathbf{\mathfrak{H}}_{\mathrm{in}}^{\left( 0\right) }\left( \mathbf{r}%
\right) &=&\frac{q}{4\pi R^{3}}\left\{ -\mathfrak{L}_{\mathfrak{G}}\mathbf{r}%
-\left[ \mathfrak{L}_{\mathfrak{FF}}\left( \overline{\mathbf{E}}\cdot 
\mathbf{r}\right) +\mathfrak{L}_{\mathfrak{FG}}\left( \mathbf{\overline{B}}%
\cdot \mathbf{r}\right) \right] \mathbf{\overline{B}}\right.  \notag \\
&+&\left. \left[ \mathfrak{L}_{\mathfrak{FG}}\left( \overline{\mathbf{E}}%
\cdot \mathbf{r}\right) +\mathfrak{L}_{\mathfrak{GG}}\left( \mathbf{%
\overline{B}}\cdot \mathbf{r}\right) \right] \overline{\mathbf{E}}\right\}
\,,\ \ r<R\,,  \label{Hin} \\
\mathbf{\mathfrak{H}}_{\mathrm{out}}^{\left( 0\right) }\left( \mathbf{r}%
\right) &=&\frac{q}{4\pi r^{3}}\left\{ -\mathfrak{L}_{\mathfrak{G}}\mathbf{r}%
-\left[ \mathfrak{L}_{\mathfrak{FF}}\left( \overline{\mathbf{E}}\cdot 
\mathbf{r}\right) +\mathfrak{L}_{\mathfrak{FG}}\left( \mathbf{\overline{B}}%
\cdot \mathbf{r}\right) \right] \mathbf{\overline{B}}\right.  \notag \\
&+&\left. \left[ \mathfrak{L}_{\mathfrak{FG}}\left( \overline{\mathbf{E}}%
\cdot \mathbf{r}\right) +\mathfrak{L}_{\mathfrak{GG}}\left( \mathbf{%
\overline{B}}\cdot \mathbf{r}\right) \right] \overline{\mathbf{E}}\right\}
\,,\ \ r\geq R\,.  \label{Hout}
\end{eqnarray}%
Here the space- and time-independent electric and magnetic components of the
background field are barred: $\overline{E}^{i}=\overline{F}_{0i},$ $%
\overline{B}^{i}=-\left( 1/2\right) \varepsilon _{ijk}\overline{F}^{jk}$.

Consider the linearly induced current density (\ref{s1.51}) to the same
first-order approximation. According to Eq. (\ref{nnew7b}), it is%
\begin{equation}
\mathbf{j}^{\left( 1\right) }\left( \mathbf{r}\right) =\boldsymbol{\nabla }%
\times \mathbf{\mathfrak{H}}^{(0)}\left( \mathbf{r}\right) =\theta (R-r)[%
\boldsymbol{\nabla }\times \mathbf{\mathfrak{H}}_{\mathrm{in}}^{(0)}\left( 
\mathbf{r}\right) ]+\theta (r-R)[\boldsymbol{\nabla }\times \mathbf{%
\mathfrak{H}}_{\mathrm{out}}^{(0)}\left( \mathbf{r}\right) ]\,.  \label{cur}
\end{equation}%
Note that the quantity (\ref{nnew13b}) is continuous at the border of the
charge $r=R,$ because $\mathbf{E}^{(0)}\left( \mathbf{r}\right) $ (\ref{in2}%
) is. For this reason the differentiation of the step functions has not
contributed to the sum (\ref{cur}). Hence there does not appear any current
at the surface of the charge.

Thus we may define the inner $\mathbf{B}_{\mathrm{in}}^{\left( 1\right)
}\left( \mathbf{r}\right) $ and outer $\mathbf{B}_{\mathrm{out}}^{\left(
1\right) }\left( \mathbf{r}\right) $ magnetic responses as solutions to the
equations%
\begin{equation}
\boldsymbol{\nabla }\times \mathbf{B}_{\mathrm{in}}^{\left( 1\right) }\left( 
\mathbf{r}\right) =\mathbf{j}_{\mathrm{in}}^{\left( 1\right) }\left( \mathbf{%
r}\right) \,,\ \ \boldsymbol{\nabla }\times \mathbf{B}_{\mathrm{out}%
}^{\left( 1\right) }\left( \mathbf{r}\right) =\mathbf{j}_{\mathrm{out}%
}^{\left( 1\right) }\left( \mathbf{r}\right) \,,  \label{B1eqs}
\end{equation}%
where the inner and outer parts of the first-order linearly induced current
densities are%
\begin{equation}
\mathbf{j}_{\mathrm{in}}^{\left( 1\right) }\left( \mathbf{r}\right) =%
\boldsymbol{\nabla }\times \mathbf{\mathfrak{H}}_{\mathrm{in}}^{(0)}\left( 
\mathbf{r}\right) =\frac{q}{4\pi R^{3}}\left( \mathfrak{L}_{\mathfrak{FF}}+%
\mathfrak{L}_{\mathfrak{GG}}\right) \left[ \mathbf{\overline{B}}\times 
\overline{\mathbf{E}}\right] \,,\ \ r<R\,,  \label{nlin}
\end{equation}%
and%
\begin{eqnarray}
\mathbf{j}_{\mathrm{out}}^{\left( 1\right) }\left( \mathbf{r}\right) &=&%
\boldsymbol{\nabla }\times \mathbf{\mathfrak{H}}_{\mathrm{out}}^{(0)}\left( 
\mathbf{r}\right) =\frac{q}{4\pi r^{3}}\left\{ \mathfrak{L}_{\mathfrak{FF}%
}\left( \left[ \overline{\mathbf{B}}\times \overline{\mathbf{E}}\right] +%
\frac{3}{r^{2}}\left( \overline{\mathbf{E}}\cdot \mathbf{r}\right) \left[ 
\mathbf{r}\times \overline{\mathbf{B}}\right] \right) \right.  \notag \\
&+&\mathfrak{L}_{\mathfrak{GG}}\left( \left[ \overline{\mathbf{B}}\times 
\overline{\mathbf{E}}\right] -\frac{3}{r^{2}}\left( \mathbf{\overline{B}}%
\cdot \mathbf{r}\right) \left[ \mathbf{r}\times \overline{\mathbf{E}}\right]
\right)  \notag \\
&+&\left. \frac{3}{r^{2}}\mathfrak{L}_{\mathfrak{FG}}\left( \left( \mathbf{%
\overline{B}}\cdot \mathbf{r}\right) \left[ \mathbf{r}\times \overline{%
\mathbf{B}}\right] -\left( \overline{\mathbf{E}}\cdot \mathbf{r}\right) %
\left[ \mathbf{r}\times \overline{\mathbf{E}}\right] \right) \right\} \,,\ \
r\geq R\,.  \label{nlout}
\end{eqnarray}%
respectively.\textrm{\ }The induced current (\ref{cur}) is discontinuous at
the edge of the sphere, the same as the charge density (\ref{const charge})
is.

For the special case of parallel external backgrounds $\overline{\mathbf{B}}%
\parallel \overline{\mathbf{E}}$, the induced current density inside the
charge disappears, $\mathbf{j}_{\mathrm{in}}^{\left( 1\right) }=0$,\ while
the current $\mathbf{j}_{\mathrm{out}}^{\left( 1\right) }\left( \mathbf{r}%
\right) $ circles the coordinate axis parallel to their common direction.
Letting $\mathbf{\hat{z}}$ be the unit vector ($\left\vert \mathbf{\hat{z}}%
\right\vert =1$) along the common direction of the external fields, $\mathbf{%
\hat{z}}\parallel \overline{\mathbf{B}}\parallel \overline{\mathbf{E}}$, the
current density $\mathbf{j}_{\mathrm{out}}^{\left( 1\right) }\left( \mathbf{r%
}\right) $ acquires the form%
\begin{equation}
\mathbf{j}_{\mathrm{out}}^{\left( 1\right) }\left( \mathbf{r}\right) =\frac{%
3q}{4\pi r^{5}}\tilde{g}\left( \mathbf{\hat{z}}\cdot \mathbf{r}\right) \left[
\mathbf{r}\times \mathbf{\hat{z}}\right] \,,  \label{par}
\end{equation}%
where $\tilde{g}$ is a combination of derivatives of the effective
Lagrangian and field invariants,%
\begin{eqnarray}
&&\tilde{g}=\overline{\mathfrak{G}}\left( \mathfrak{L}_{\mathfrak{GG}}-%
\mathfrak{L}_{\mathfrak{FF}}\right) +2\overline{\mathfrak{F}}\mathfrak{L}_{%
\mathfrak{FG}}\,,  \notag \\
&&\overline{\mathfrak{G}}=-\mathbf{\overline{B}}\cdot \overline{\mathbf{E}}%
\,,\ \ \overline{\mathfrak{F}}=\frac{1}{2}\left( \mathbf{\overline{B}}^{2}-%
\overline{\mathbf{E}}^{2}\right) \,.  \label{gtilded}
\end{eqnarray}

\begin{figure}[th]
\begin{center}
\includegraphics[scale=1.0]{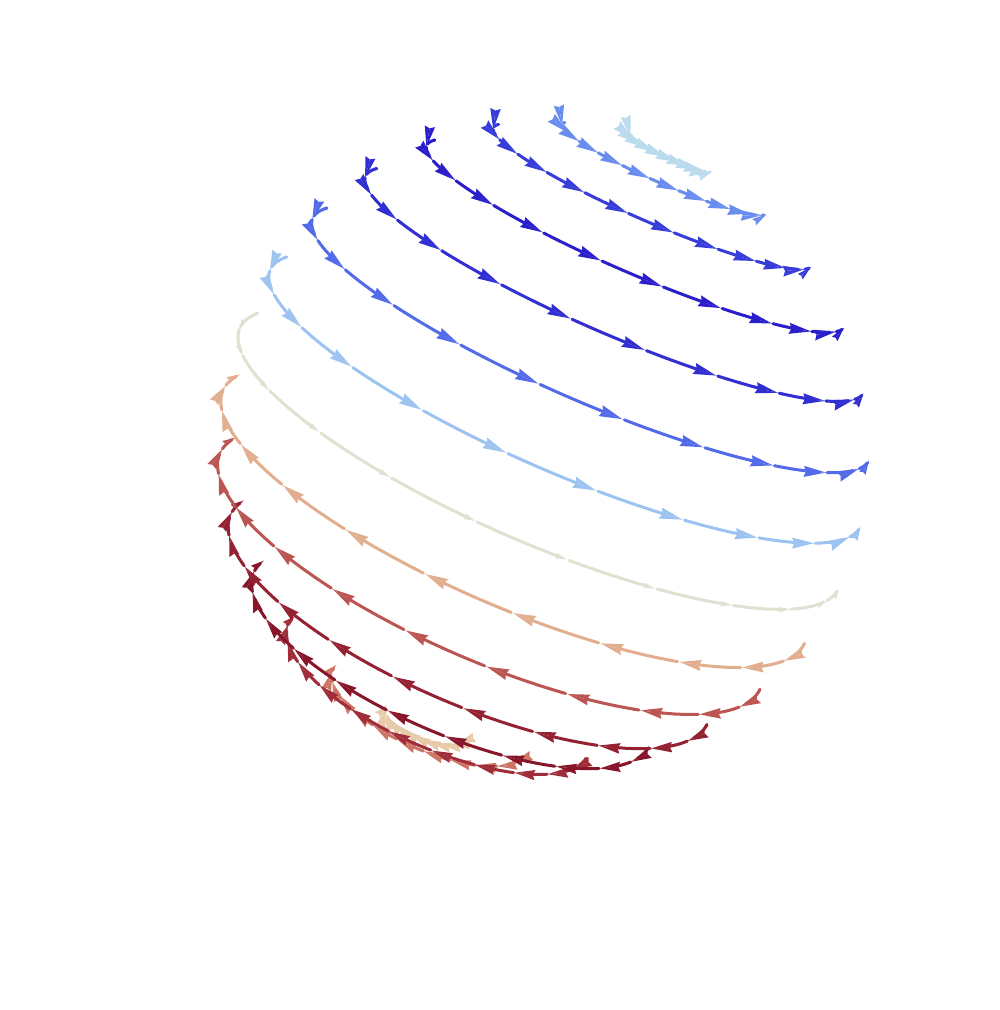}
\end{center}
\caption{(color online) Flux of the electric current (\protect\ref{par})
linearly induced by static electric charge placed into parallel electric and
magnetic background fields (outside the charge). The current revolves about
the axis drawn along the common direction of the background fields. The
brighter the arrow-head lines, the larger the current density.}
\label{Fig1}
\end{figure}

The current flux (\ref{par}) flows in opposite directions in the upper and
lower hemispheres; see Fig. \ref{Fig1}. Hence, the total current through the
part of a fixed meridial plane $\varphi =\varphi _{0},$\ $0<\varphi
_{0}<2\pi ,$\ enclosed between any two coordinate spheres $r_{1}<r<r_{2}$\
is zero:%
\begin{equation*}
\int \mathbf{j}_{\mathrm{out}}^{\left( 1\right) }\left( \mathbf{r}\right) d%
\mathbf{s}=\frac{3q}{4\pi }\tilde{g}\int_{r_{1}}^{r_{2}}\frac{dr}{r^{2}}%
\int_{0}^{\pi }d\theta \cos \theta \sin \theta =0\,.
\end{equation*}%
\ Here $\cos \theta =\left( \mathbf{\hat{z}}\cdot \mathbf{r}\right) /r$.\
Once any two mutually non-orthogonal constant fields may be reduced to
parallelity by an approporiate Lorentz transformation to a special inertial
frame, the current (\ref{nlout}) differs from (\ref{par}) by a contribution
due to the motion of the charge in that frame. 

\subsection{Magnetic response\label{Ss3.1}}

Besides the linearized Maxwell equations (\ref{nnew7b}), the magnetic
response $\mathbf{B}^{\left( 1\right) }\left( \mathbf{r}\right) $ should
obey also the equation%
\begin{equation}
\boldsymbol{\nabla }\cdot \mathbf{B}^{\left( 1\right) }\left( \mathbf{r}%
\right) =0\,,  \label{bianchi}
\end{equation}%
\ that excludes an overall magnetic charge and makes the formulation of the
theory in terms of potentials possible, in which case it corresponds to one
of the Bianchi identities in electrodynamics. Equations (\ref{nnew7b}) and (%
\ref{bianchi}) are satisfied by the magnetic response\ $\mathbf{B}^{\left(
1\right) }\left( \mathbf{r}\right) $%
\begin{equation}
B^{\left( 1\right) i}\left( \mathbf{r}\right) =\left( \delta ^{ij}-\frac{%
\partial _{i}\partial _{j}}{\mathbf{\nabla }^{2}}\right) \mathfrak{H}%
^{\left( 0\right) j}\left( \mathbf{r}\right) \,,  \label{sc2.5}
\end{equation}%
identified as the transverse component of $\mathbf{\mathfrak{H}}^{\left(
0\right) }\left( \mathbf{r}\right) $. The integral form of (\ref{sc2.5})
reads%
\begin{equation}
B^{\left( 1\right) i}\left( \mathbf{r}\right) =\mathfrak{H}^{\left( 0\right)
i}\left( \mathbf{r}\right) +\frac{1}{4\pi }\partial _{i}\partial _{j}\int d%
\mathbf{y}\frac{\mathfrak{H}^{\left( 0\right) j}\left( \mathbf{y}\right) }{%
\left\vert \mathbf{r-y}\right\vert }\,.  \label{sc2.5b}
\end{equation}%
After computing these integrals (see Eqs. (\ref{apa0}) - (\ref{arb19}) in
Appendix \ref{ApA}) the inner part $\mathbf{B}_{\mathrm{in}}^{\left(
1\right) }\left( \mathbf{r}\right) $ takes the form%
\begin{eqnarray}
\mathbf{B}_{\mathrm{in}}^{\left( 1\right) }\left( \mathbf{r}\right)  &=&%
\frac{q}{4\pi }\left\{ \left[ \frac{3}{5R^{3}}\mathfrak{L}_{\mathfrak{FG}%
}\left( \overline{\mathbf{E}}\cdot \mathbf{r}\right) +\frac{1}{5R^{3}}\left( 
\mathfrak{L}_{\mathfrak{FF}}+4\mathfrak{L}_{\mathfrak{GG}}\right) \left( 
\mathbf{\overline{B}}\cdot \mathbf{r}\right) \right] \overline{\mathbf{E}}%
\right.   \notag \\
&+&\left. \left[ -\frac{3}{5R^{3}}\mathfrak{L}_{\mathfrak{FG}}\left( \mathbf{%
\overline{B}}\cdot \mathbf{r}\right) -\frac{1}{5R^{3}}\left( 4\mathfrak{L}_{%
\mathfrak{FF}}+\mathfrak{L}_{\mathfrak{GG}}\right) \left( \overline{\mathbf{E%
}}\cdot \mathbf{r}\right) \right] \mathbf{\overline{B}}+\frac{\tilde{g}}{%
5R^{3}}\mathbf{r}\right\} \,,  \label{arb21}
\end{eqnarray}%
while the outer part $\mathbf{B}_{\mathrm{out}}^{\left( 1\right) }\left( 
\mathbf{r}\right) $ can be conveniently written as%
\begin{equation}
\mathbf{B}_{\mathrm{out}}^{\left( 1\right) }\left( \mathbf{r}\right) =%
\mathbf{B}_{\mathrm{pl}}^{\left( 1\right) }\left( \mathbf{r}\right) +\mathbf{%
B}_{\mathrm{out}}^{\left( 1\right) }\left( \mathbf{r};R\right) \,.
\label{arb22}
\end{equation}%
Here $\mathbf{B}_{\mathrm{pl}}^{\left( 1\right) }\left( \mathbf{r}\right) $
denotes an $R$-free part%
\begin{eqnarray}
\mathbf{B}_{\mathrm{pl}}^{\left( 1\right) }\left( \mathbf{r}\right)  &=&%
\frac{q}{4\pi }\left\{ \frac{\tilde{g}}{2}\frac{\mathbf{r}}{r^{3}}-\frac{%
\left( \mathfrak{L}_{\mathfrak{GG}}+\mathfrak{L}_{\mathfrak{FF}}\right) }{%
2r^{3}}\left[ \left( \overline{\mathbf{E}}\cdot \mathbf{r}\right) \mathbf{%
\overline{B}}-\left( \mathbf{\overline{B}}\cdot \mathbf{r}\right) \overline{%
\mathbf{E}}\right] \right.   \notag \\
&+&\left. \frac{3}{2}\left[ \left( \mathfrak{L}_{\mathfrak{GG}}-\mathfrak{L}%
_{\mathfrak{FF}}\right) \left( \overline{\mathbf{E}}\cdot \mathbf{r}\right)
\left( \mathbf{\overline{B}}\cdot \mathbf{r}\right) -\mathfrak{L}_{\mathfrak{%
FG}}\left( \left( \mathbf{\overline{B}}\cdot \mathbf{r}\right) ^{2}-\left( 
\overline{\mathbf{E}}\cdot \mathbf{r}\right) ^{2}\right) \right] \frac{%
\mathbf{r}}{r^{5}}\right\} \,,  \label{arb220}
\end{eqnarray}%
while $\mathbf{B}_{\mathrm{out}}^{\left( 1\right) }\left( \mathbf{r}%
;R\right) $, the $R$-dependent part, reads%
\begin{eqnarray}
\mathbf{B}_{\mathrm{out}}^{\left( 1\right) }\left( \mathbf{r};R\right)  &=&%
\frac{q}{8\pi }\left( \frac{3R^{2}}{5r^{5}}\right) \left\{ 2\mathfrak{L}_{%
\mathfrak{FG}}\left[ \left( \overline{\mathbf{E}}\cdot \mathbf{r}\right) 
\overline{\mathbf{E}}-\left( \mathbf{\overline{B}}\cdot \mathbf{r}\right) 
\mathbf{\overline{B}}\right] \right.   \notag \\
&-&\left( \mathfrak{L}_{\mathfrak{FF}}-\mathfrak{L}_{\mathfrak{GG}}\right) %
\left[ \left( \mathbf{\overline{B}}\cdot \mathbf{r}\right) \overline{\mathbf{%
E}}+\left( \overline{\mathbf{E}}\cdot \mathbf{r}\right) \mathbf{\overline{B}}%
\right]   \notag \\
&+&\left[ -\tilde{g}+\frac{5}{r^{2}}\left( \mathfrak{L}_{\mathfrak{FF}}-%
\mathfrak{L}_{\mathfrak{GG}}\right) \left( \overline{\mathbf{E}}\cdot 
\mathbf{r}\right) \left( \mathbf{\overline{B}}\cdot \mathbf{r}\right)
\right.   \notag \\
&+&\left. \left. \frac{5}{r^{2}}\mathfrak{L}_{\mathfrak{FG}}\left( \left( 
\mathbf{\overline{B}}\cdot \mathbf{r}\right) ^{2}-\left( \overline{\mathbf{E}%
}\cdot \mathbf{r}\right) ^{2}\right) \right] \mathbf{r}\right\} \,,
\label{arb22R}
\end{eqnarray}%
The division in $R$-dependent and $R$-free terms, expressed in Eq. (\ref%
{arb22}), is aimed to emphasize that Eq. (\ref{arb22R}) corresponds to a
pure homogeneous solution $\boldsymbol{\nabla }\times \mathbf{B}_{\mathrm{out%
}}^{\left( 1\right) }\left( \mathbf{r};R\right) =0$. This is a consequence
of the fact that the outer induced current density, given by Eq. (\ref{nlout}%
), does not depend on $R$ or, in other words, there is no $R$-dependent
source providing (\ref{arb22R}). Its real role is to provide continuity of
the whole magnetic response $\mathbf{B}^{\left( 1\right) }\left( \mathbf{r}%
\right) =\mathbf{B}_{\mathrm{in}}^{\left( 1\right) }\left( \mathbf{r}\right)
\theta \left( R-r\right) +\mathbf{B}_{\mathrm{out}}^{\left( 1\right) }\left( 
\mathbf{r}\right) \theta \left( r-R\right) $ at the border of the Coulomb
source (\ref{const charge}). A similar feature has been reported by us in 
\cite{AdoGitSha2016}, wherein $R$-dependent terms in the electric response
come automatically from the projection operator with the same
interpretation. These $R$-dependent solutions are a consequence of the
Coulomb source being an extended charge distribution rather than a pointlike
one. In contrast, the\ $R$-independent part $\mathbf{B}_{\mathrm{pl}%
}^{\left( 1\right) }\left( \mathbf{r}\right) $ is the same as the
first-order linear response of the pointlike Coulomb source (\ref{pointlike}%
), since\ it is the only surviver in the limit $r\gg R$ (or $R\rightarrow 0$%
), bearing in mind that for any nonspecial direction, $\mathbf{B}_{\mathrm{pl%
}}^{\left( 1\right) }\left( \mathbf{r}\right) $ decreases as $r^{-2}$,\
while $\mathbf{B}_{\mathrm{out}}^{\left( 1\right) }\left( \mathbf{r}%
;R\right) $ decreases as $r^{-4}$. Therefore $\mathbf{B}_{\mathrm{pl}%
}^{\left( 1\right) }\left( \mathbf{r}\right) $ is identified as the
first-order linear response to the pointlike Coulomb source (\ref{pointlike}%
). Moreover, according to Eq. (\ref{nlout}), $\mathbf{B}_{\mathrm{pl}%
}^{\left( 1\right) }\left( \mathbf{r}\right) $ is provided by the outer
induced current $\mathbf{j}_{\mathrm{out}}^{\left( 1\right) }\left( \mathbf{r%
}\right) $%
\begin{equation}
\boldsymbol{\nabla }\times \mathbf{B}_{\mathrm{pl}}^{\left( 1\right) }\left( 
\mathbf{r}\right) =\boldsymbol{\nabla }\times \mathbf{\mathfrak{H}}_{\mathrm{%
out}}^{\left( 0\right) }\left( \mathbf{r}\right) =\mathbf{j}_{\mathrm{out}%
}^{\left( 1\right) }\left( \mathbf{r}\right) \,.  \label{arb23}
\end{equation}

The first-order linear magnetic response calculated above does not carry any
magnetic charge, in virtue of the triviality of the Gauss integral%
\begin{equation}
\oint_{S}\left( \mathbf{B}^{\left( 1\right) }\left( \mathbf{r}\right) \cdot 
\mathbf{\hat{n}}\right) dS=0\,,  \label{arb230}
\end{equation}%
for an arbitrary closed surface $S$ embracing the charge $q$. This integral
vanishes for each magnetic response $\mathbf{B}_{\mathrm{in}}^{\left(
1\right) }\left( \mathbf{r}\right) $, $\mathbf{B}_{\mathrm{pl}}^{\left(
1\right) }\left( \mathbf{r}\right) $ and $\mathbf{B}_{\mathrm{out}}^{\left(
1\right) }\left( \mathbf{r};R\right) $, independently. To begin with, taking 
$S$ to be a sphere of radius $R$, centered in the charge $q$ (placed at the
origin $r=0$), and choosing a reference frame in which $\overline{\mathbf{B}}
$ is aligned along the $z$-axis and $\overline{\mathbf{E}}$ lies in the $xz$%
-plane (so that $\overline{\mathbf{E}}\cdot \mathbf{r}=\overline{E}R\cos
\gamma $\thinspace ,\ $\cos \gamma =\cos \theta \cos \theta ^{\prime }+\sin
\theta \sin \theta ^{\prime }\cos \varphi $ and $\overline{\mathbf{E}}\cdot 
\overline{\mathbf{B}}=\overline{E}\overline{B}\cos \theta ^{\prime }$), we
find that only $\mathbf{B}_{\mathrm{in}}^{\left( 1\right) }\left( \mathbf{r}%
\right) $ might contribute in Eq. (\ref{arb230}) and that%
\begin{equation}
\oint_{S}\left( \mathbf{B}_{\mathrm{in}}^{\left( 1\right) }\left( \mathbf{r}%
\right) \cdot \mathbf{\hat{n}}\right) dS=\frac{3q}{20\pi }\left( -\frac{%
\mathcal{J}\left( R\right) }{R^{2}}+\frac{4\pi \tilde{g}}{3}\right) =0\,,
\label{arb231}
\end{equation}%
where $\mathcal{J}\left( R\right) $ denotes the integral over the surface $S$%
\begin{eqnarray}
\mathcal{J}\left( R\right) &=&\int_{0}^{2\pi }d\varphi \int_{0}^{\pi
}d\theta \sin \theta \left\{ \left( \mathfrak{L}_{\mathfrak{FF}}-\mathfrak{L}%
_{\mathfrak{GG}}\right) \left( \overline{\mathbf{E}}\cdot \mathbf{r}\right)
\left( \mathbf{\overline{B}}\cdot \mathbf{r}\right) \right.  \notag \\
&+&\left. \mathfrak{L}_{\mathfrak{FG}}\left[ \left( \mathbf{\overline{B}}%
\cdot \mathbf{r}\right) ^{2}-\left( \overline{\mathbf{E}}\cdot \mathbf{r}%
\right) ^{2}\right] \right\} =\frac{4\pi R^{2}}{3}\tilde{g}\,.  \label{iden}
\end{eqnarray}%
Therefore Eq. (\ref{arb231}) holds true. If one takes $S$ to be a sphere of
radius $r>R$, then Eq. (\ref{arb230}) takes the form%
\begin{equation*}
\oint_{S}\left( \mathbf{B}_{\mathrm{out}}^{\left( 1\right) }\left( \mathbf{r}%
\right) \cdot \mathbf{\hat{n}}\right) dS=\oint_{S}\left( \mathbf{B}_{\mathrm{%
pl}}^{\left( 1\right) }\left( \mathbf{r}\right) \cdot \mathbf{\hat{n}}%
\right) dS+\oint_{S}\left( \mathbf{B}_{\mathrm{out}}^{\left( 1\right)
}\left( \mathbf{r};R\right) \cdot \mathbf{\hat{n}}\right) dS\,,
\end{equation*}%
in which it can be seen that both integrals vanish identically%
\begin{eqnarray}
\oint_{S}\left( \mathbf{B}_{\mathrm{out}}^{\left( 1\right) }\left( \mathbf{r}%
;R\right) \cdot \mathbf{\hat{n}}\right) dS &=&-\frac{3R^{2}}{5r^{2}}%
\oint_{S}\left( \mathbf{B}_{\mathrm{pl}}^{\left( 1\right) }\left( \mathbf{r}%
\right) \cdot \mathbf{\hat{n}}\right) dS  \notag \\
&=&-\frac{3R^{2}}{5r^{2}}\frac{q}{8\pi }\left( 4\pi \tilde{g}-\frac{3%
\mathcal{J}\left( r\right) }{r^{2}}\right) =0\,.  \label{arb232}
\end{eqnarray}%
Here $\mathcal{J}\left( r\right) $ is the same integral as (\ref{iden}), but
with the $R$ replaced by $r$. One concludes that there is no magnetic charge
attributed to the magnetic response $\mathbf{B}^{\left( 1\right) }\left( 
\mathbf{r}\right) $: the magnetic lines of force incomming to and ougoing
from the charge $q$ , compensate each other, so that the corresponding
magnetic flux be zero.

\begin{figure}[th]
\begin{center}
\includegraphics[scale=0.43]{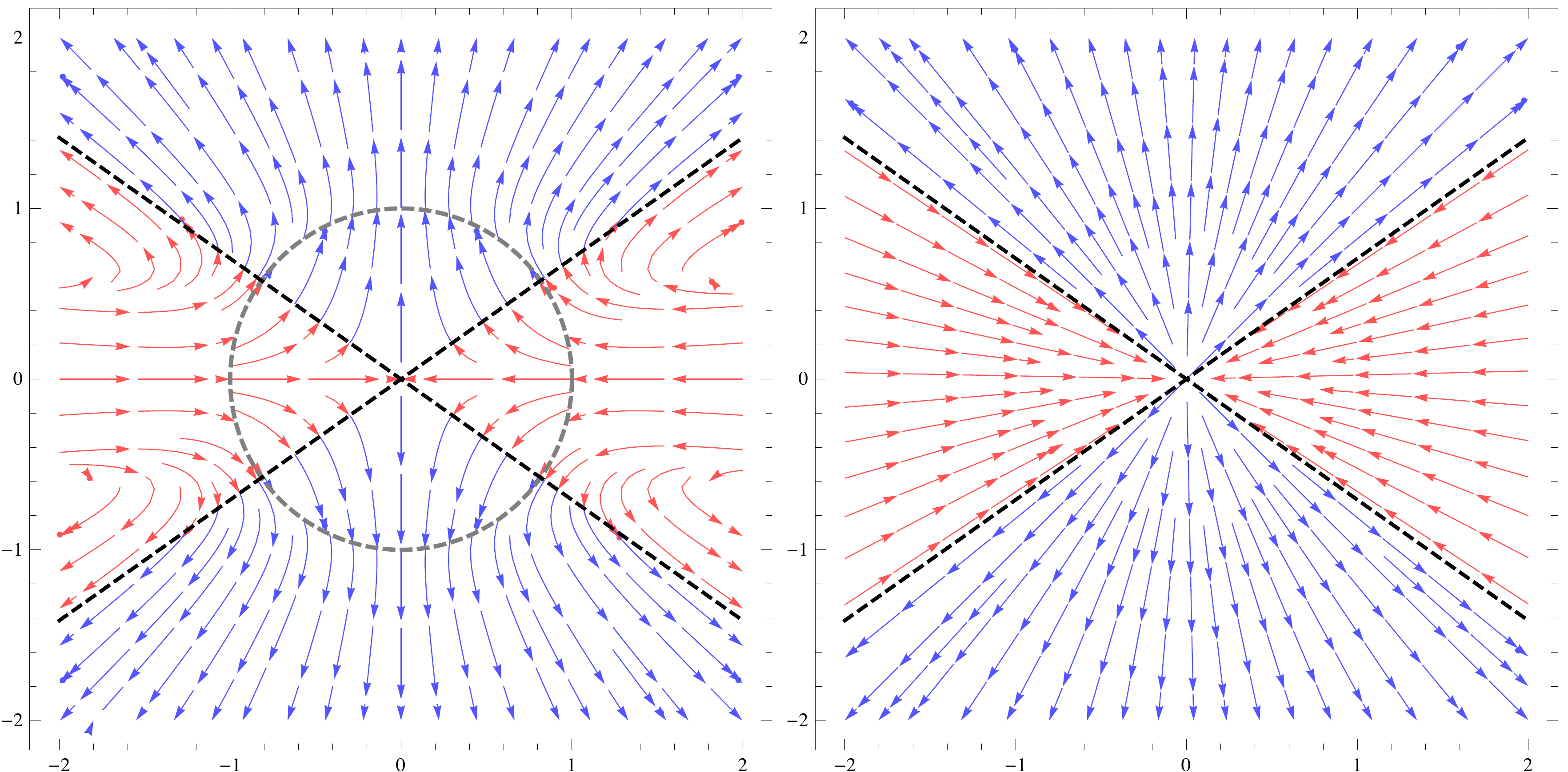}
\end{center}
\caption{(color online) Magnetic lines of force produced in a constant
background by the static charge. The left pattern corresponds to the
extended charge concentrated within the dashed circle, and the right pattern
corresponds to the point charge (according to Eq. (\protect\ref{ansa20})).
Magnetic field $\mathbf{B}_{\mathrm{in}}^{\left( 1\right) }\left( \mathbf{x}%
\right) $ inside the charge is drawn following Eq. (\protect\ref{arb23.1}),
and its outer part $\mathbf{B}_{\mathrm{out}}^{\left( 1\right) }\left( 
\mathbf{x}\right) $ following Eq. (\protect\ref{arb23.2}). In the limit $%
R\rightarrow 0$, the red/blue arrows in the left pattern tend to straight
lines, becoming \textquotedblleft inward\textquotedblright\ and
\textquotedblleft outward\textquotedblright\ magnetic lines of force, as
depicted in the right pattern. The inclined dashed lines indicate regions of
zero magnetic field on the right, whereas, on the left, they only divide
\textquotedblleft inward\textquotedblright\ and \textquotedblleft
outward\textquotedblright\ magnetic lines of force in the limit $%
R\rightarrow 0$.}
\label{Fig2}
\end{figure}

To visualize the structure of the magnetic lines of force, let us consider
the particular case of parallel background fields, $\overline{\mathbf{E}}=%
\overline{E}\mathbf{\hat{z}}\,,\ \ \mathbf{\overline{B}}=\overline{B}\mathbf{%
\hat{z}}$ ($\left\vert \mathbf{\hat{z}}\right\vert =1$), whose response
acquires a simpler form%
\begin{equation}
\mathbf{B}_{\mathrm{in}}^{\left( 1\right) }\left( \mathbf{r}\right) =\frac{q%
\tilde{g}}{4\pi R^{3}}\frac{1}{5}\left[ \mathbf{r}-3\left( \mathbf{\hat{z}}%
\cdot \mathbf{r}\right) \mathbf{\hat{z}}\right] \,,  \label{arb23.1}
\end{equation}%
and%
\begin{eqnarray}
\mathbf{B}_{\mathrm{out}}^{\left( 1\right) }\left( \mathbf{r}\right) &=&%
\frac{q\tilde{g}}{4\pi r^{3}}\left\{ \left[ 1-\frac{3R^{2}}{5r^{2}}-3\left(
1-\frac{R^{2}}{r^{2}}\right) \left( \frac{\mathbf{\hat{z}}\cdot \mathbf{r}}{r%
}\right) ^{2}\right] \frac{\mathbf{r}}{2}\right.  \notag \\
&-&\left. \frac{3R^{2}}{5r^{2}}\left( \mathbf{\hat{z}}\cdot \mathbf{r}%
\right) \mathbf{\hat{z}}\right\} \,.  \label{arb23.2}
\end{eqnarray}%
Moreover, in the limit $r/R\rightarrow \infty $, we are left with a single
magnetic response, exclusively radial, although spherically nonsymmetric,
corresponding to that of a pointlike Coulomb source (\ref{pointlike})%
\begin{equation}
\mathbf{B}_{\mathrm{pl}}^{\left( 1\right) }\left( \mathbf{r}\right)
=\lim_{r/R\rightarrow \infty }\mathbf{B}^{\left( 1\right) }\left( \mathbf{r}%
\right) =\frac{q}{4\pi }\frac{\tilde{g}}{2}\left[ 1-3\left( \frac{\mathbf{%
\hat{z}}\cdot \mathbf{r}}{r}\right) ^{2}\right] \frac{\mathbf{r}}{r^{3}}\,.
\label{ansa20}
\end{equation}%
The magnetic lines of force are straight lines, vanishing at the angles $%
\cos \theta =\zeta =\zeta _{0}=1/\sqrt{3}$. As no net magnetic charge exists
for producing a nontrivial magnetic flux (\ref{arb232}), there are inward
magnetic lines (pointing to $q$) and outward magnetic lines (lines leaving $%
q $), in the same proportion (see Fig \ref{Fig2}).

The magnetic field found (\ref{ansa20}) may be understood as two pointlike
magnetic poles of equal, but opposite, polarities superposed in one point,
so to say, a pointlike magnetic dipole.

\section{Vector potentials\label{S4}}

In this section, we extend the consideration above to the level of
electromagnetic potentials, restricting ourselves to the case of
electromagnetic responses generated by a pointlike charge distribution (\ref%
{pointlike}) since, as discussed before, the role of the regularization (\ref%
{const charge}) is simply to avoid divergent integrals in the calculation of
magnetic responses (\ref{sc2.5b}) if a pointlike source is considered from
the beginning. In this case, the vector potential is sought in the form%
\footnote{%
For the sake of convenience, we remove the subscript \textquotedblleft 
\textrm{pl}\textquotedblright\ on the magnetic response generated by the
pointlike Coulomb distribution (\ref{pointlike}), given by Eq. (\ref{arb220}%
). Thus $\mathbf{B}_{\mathrm{pl}}^{\left( 1\right) }\left( \mathbf{r}\right)
\equiv \mathbf{B}^{\left( 1\right) }\left( \mathbf{r}\right) $ from now on.},%
\begin{eqnarray}
&&\mathbf{A}^{\left( 1\right) }\left( \mathbf{r}\right) =\left[ \mathbf{\hat{%
z}}\times \mathbf{r}\right] \frac{\mathcal{A}\left( \zeta ,\xi \right) }{%
r^{2}}+\left[ \boldsymbol{\hat{\nu}}\times \mathbf{r}\right] \frac{\mathcal{C%
}\left( \zeta ,\xi \right) }{r^{2}}+\frac{\left[ \mathbf{\hat{z}}\times 
\boldsymbol{\hat{\nu}}\right] }{r}\mathcal{M}\,,  \notag \\
&&\mathbf{\hat{z}}=\frac{\overline{\mathbf{B}}}{\overline{B}}\,,\ \ 
\boldsymbol{\hat{\nu}}=\frac{\overline{\mathbf{E}}}{\overline{E}}\,,\ \
\zeta =\frac{\mathbf{\hat{z}}\cdot \mathbf{r}}{r}\,,\ \ \xi =\frac{%
\boldsymbol{\hat{\nu}}\cdot \mathbf{r}}{r}\,,  \label{4.1}
\end{eqnarray}%
where $\mathcal{M}$ is a constant and $\mathcal{A}\left( \zeta ,\xi \right) $%
, $\mathcal{C}\left( \zeta ,\xi \right) $ are functions of the cosines of
the angles between directions of $\overline{\mathbf{E}}$, $\overline{\mathbf{%
B}}$ and the radius vector $\mathbf{r}$. Although this form may be not the
most general one, its use is sufficient for finding at least a certain class
of the vector-potentials $\mathbf{A}^{\left( 1\right) }\left( \mathbf{r}%
\right) $, all the variety of other possible values for $\mathbf{A}^{\left(
1\right) }\left( \mathbf{r}\right) $\ being gauge-equivalent to those found.
The representation of the field in terms of the vector potential should be
exploited I changed this phrase a little, Shabad:%
\begin{eqnarray}
\mathbf{B}^{\left( 1\right) }\left( \mathbf{r}\right)  &=&\boldsymbol{\nabla 
}\times \mathbf{A}^{\left( 1\right) }\left( \mathbf{r}\right) =\frac{\mathbf{%
r}}{r^{4}}\left[ 2\mathcal{A}\left( \mathbf{\hat{z}}\cdot \mathbf{r}\right)
+2\mathcal{C}\left( \boldsymbol{\hat{\nu}}\cdot \mathbf{r}\right) \right] -%
\frac{\mathbf{r}}{r^{3}}\left( \mathbf{\hat{z}}\cdot \boldsymbol{\hat{\nu}}-%
\frac{\left( \mathbf{\hat{z}}\cdot \mathbf{r}\right) \left( \boldsymbol{\hat{%
\nu}}\cdot \mathbf{r}\right) }{r^{2}}\right) \left( \partial _{\xi }\mathcal{%
A}+\partial _{\zeta }\mathcal{C}\right)   \notag \\
&-&\frac{\mathbf{r}}{r^{3}}\left\{ \left[ 1-\left( \frac{\mathbf{\hat{z}}%
\cdot \mathbf{r}}{r}\right) ^{2}\right] \partial _{\zeta }\mathcal{A}+\left[
1-\left( \frac{\boldsymbol{\hat{\nu}}\cdot \mathbf{r}}{r}\right) ^{2}\right]
\partial _{\xi }\mathcal{C}\right\} +\frac{\mathcal{M}}{r^{3}}\left[ 
\boldsymbol{\hat{\nu}}\left( \mathbf{\hat{z}}\cdot \mathbf{r}\right) -%
\mathbf{\hat{z}}\left( \boldsymbol{\hat{\nu}}\cdot \mathbf{r}\right) \right]
\,.  \label{H}
\end{eqnarray}%
Note that only the $\mathcal{M}$-term from (\ref{4.1}) contributes the $%
\mathbf{\hat{z}}$ - and $\boldsymbol{\hat{\nu}}$ - components to (\ref{H}).
The two equations obtained by projecting the equation resulting from
equating (\ref{arb220}) and (\ref{H}) onto these directions%
\begin{equation*}
-\frac{q}{4\pi }\frac{\left( \mathfrak{L}_{\mathfrak{GG}}+\mathfrak{L}_{%
\mathfrak{FF}}\right) }{2r^{3}}\left[ \left( \overline{\mathbf{E}}\cdot 
\mathbf{r}\right) \mathbf{\overline{B}}-\left( \mathbf{\overline{B}}\cdot 
\mathbf{r}\right) \overline{\mathbf{E}}\right] =-\frac{\mathcal{M}}{r^{2}}%
\left[ \mathbf{\hat{z}}\left( \boldsymbol{\hat{\nu}}\cdot \mathbf{r}\right) -%
\boldsymbol{\hat{\nu}}\left( \mathbf{\hat{z}}\cdot \mathbf{r}\right) \right]
\,,
\end{equation*}%
are both satisfied with the unique choice%
\begin{equation}
\mathcal{M}=\frac{q}{8\pi }\left( \mathfrak{L}_{\mathfrak{FF}}+\mathfrak{L}_{%
\mathfrak{GG}}\right) \overline{B}\overline{E}\,.  \label{4.3}
\end{equation}%
This fact has an important consequence: after projecting (\ref{arb220}), (%
\ref{H}) onto the $\mathbf{r}$-direction we are left with only one
first-order partial differential equation for two functions $\mathcal{A}$
and $\mathcal{C}$%
\begin{eqnarray}
&&2\left( \zeta \mathcal{A}+\xi \mathcal{C}\right) -\left[ \left( 1-\zeta
^{2}\right) \partial _{\zeta }\mathcal{A}+\left( 1-\xi ^{2}\right) \partial
_{\xi }\mathcal{C}\right] -\left( \mathbf{\hat{z}}\cdot \boldsymbol{\hat{\nu}%
}-\zeta \xi \right) \left( \partial _{\xi }\mathcal{A}+\partial _{\zeta }%
\mathcal{C}\right)   \notag \\
&=&\frac{q}{4\pi }\frac{\tilde{g}}{2}+\frac{3q}{8\pi }\left[ \left( 
\mathfrak{L}_{\mathfrak{GG}}-\mathfrak{L}_{\mathfrak{FF}}\right) \overline{B}%
\overline{E}\zeta \xi -\mathfrak{L}_{\mathfrak{FG}}\overline{B}^{2}\zeta
^{2}+\mathfrak{L}_{\mathfrak{FG}}\overline{E}^{2}\xi ^{2}\right] \,.
\label{4.2}
\end{eqnarray}%
Therefore, there is an arbitrary number of solutions to Eq. (\ref{4.2}).
Such arbitrariness corresponds to a certain part of the gauge freedom of
vector potentials. It will be sufficient to seek for solutions of Eq. (\ref%
{4.2}) in the following subclass of (\ref{4.1})%
\begin{eqnarray}
&&\mathcal{A}\left( \zeta ,\xi \right) =\frac{q}{16\pi }\left( \mathfrak{L}_{%
\mathfrak{GG}}-\mathfrak{L}_{\mathfrak{FF}}\right) \overline{B}\overline{E}%
\xi -\frac{q}{8\pi }\mathfrak{L}_{\mathfrak{FG}}\overline{B}^{2}\zeta
+Y\left( \zeta \right) \,,  \notag \\
&&\mathcal{C}\left( \zeta ,\xi \right) =\frac{q}{16\pi }\left( \mathfrak{L}_{%
\mathfrak{GG}}-\mathfrak{L}_{\mathfrak{FF}}\right) \overline{B}\overline{E}%
\zeta +\frac{q}{8\pi }\mathfrak{L}_{\mathfrak{FG}}\overline{E}^{2}\xi
+X\left( \xi \right) \,,  \label{4.4}
\end{eqnarray}%
where the functions $Y\left( \zeta \right) $ and $X\left( \xi \right) $
satisfy one and the same differential equation,%
\begin{equation}
Z^{\prime }\left( u\right) -\left( \frac{2u}{1-u^{2}}\right) Z\left(
u\right) =0\,,  \label{4.5}
\end{equation}%
in which $Z=\left( Y,X\right) $ and $u=\left( \zeta ,\xi \right) $. The
latter equation can be readily integrated,%
\begin{equation}
Z\left( u\right) =\frac{z^{2}-1}{u^{2}-1}Z\left( z\right) \,,  \label{4.5.2}
\end{equation}%
yielding the final form of the vector potential (\ref{4.1}),%
\begin{eqnarray}
\mathbf{A}^{\left( 1\right) }\left( \mathbf{r}\right)  &=&\frac{\left[ 
\overline{\mathbf{B}}\times \mathbf{r}\right] }{r^{2}}\left[ \frac{q\left( 
\mathfrak{L}_{\mathfrak{GG}}-\mathfrak{L}_{\mathfrak{FF}}\right) }{16\pi }%
\frac{\overline{\mathbf{E}}\cdot \mathbf{r}}{r}-\frac{q\mathfrak{L}_{%
\mathfrak{FG}}}{8\pi }\frac{\overline{\mathbf{B}}\cdot \mathbf{r}}{r}+\frac{%
z^{2}-1}{\zeta ^{2}-1}Y\left( z\right) \right]   \notag \\
&+&\frac{\left[ \overline{\mathbf{E}}\times \mathbf{r}\right] }{r^{2}}\left[ 
\frac{q\left( \mathfrak{L}_{\mathfrak{GG}}-\mathfrak{L}_{\mathfrak{FF}%
}\right) }{16\pi }\frac{\overline{\mathbf{B}}\cdot \mathbf{r}}{r}+\frac{q%
\mathfrak{L}_{\mathfrak{FG}}}{8\pi }\frac{\overline{\mathbf{E}}\cdot \mathbf{%
r}}{r}+\frac{\tilde{z}^{2}-1}{\xi ^{2}-1}X\left( \tilde{z}\right) \right]  
\notag \\
&+&\frac{q}{8\pi }\left( \mathfrak{L}_{\mathfrak{FF}}+\mathfrak{L}_{%
\mathfrak{GG}}\right) \frac{\left[ \overline{\mathbf{B}}\times \overline{%
\mathbf{E}}\right] }{r^{2}}\,,  \label{4.6}
\end{eqnarray}%
In Eqs. (\ref{4.5.2}) and (\ref{4.6}), $z,\tilde{z}\in \left[ -1,+1\right] $
are integration constants or boundary points, through which the boundary
conditions $Y\left( z\right) $, $X\left( \tilde{z}\right) $ should be
specified. The choice of the latter is a matter of gauge fixing. It should
be noted that in the class of functions given by Eqs. (\ref{4.4}), the
boundary conditions cannot be fixed by imposing the Coulomb gauge, since%
\begin{equation}
\boldsymbol{\nabla }\cdot \mathbf{A}^{\left( 1\right) }\left( \mathbf{r}%
\right) =\frac{q\left( \mathfrak{L}_{\mathfrak{FF}}+\mathfrak{L}_{\mathfrak{%
GG}}\right) }{8\pi }\frac{\mathbf{r}\cdot \left[ \overline{\mathbf{E}}\times 
\overline{\mathbf{B}}\right] }{r^{3}}\,.  \label{gauge}
\end{equation}%
The arbitrariness due to different choices of the boundary conditions leaves
us within this gauge. Nevertheless, there are two special choices for the
integration constants, namely $z=\tilde{z}=\pm 1$, that do not require that
these boundary conditions be specified. For any other choices of $z$ and $%
\tilde{z}$, the potential is singular along the entire $z$- and $\nu $-axes,
corresponding to the direction of the external fields $\overline{\mathbf{B}}$%
, $\overline{\mathbf{E}}$, respectively, provided $Y\left( z\right) $ and $%
X\left( \tilde{z}\right) $ are nontrivial.

Sticking to the choices $z=\tilde{z}=\pm 1$, the vector potential (\ref{4.6}%
) acquires the final form%
\begin{eqnarray}
\mathbf{A}^{\left( 1\right) }\left( \mathbf{r}\right)  &=&\frac{q}{16\pi
r^{2}}\left[ \overline{\mathbf{B}}\times \mathbf{r}\right] \left[ \left( 
\mathfrak{L}_{\mathfrak{GG}}-\mathfrak{L}_{\mathfrak{FF}}\right) \left( 
\frac{\overline{\mathbf{E}}\cdot \mathbf{r}}{r}\right) -2\mathfrak{L}_{%
\mathfrak{FG}}\left( \frac{\overline{\mathbf{B}}\cdot \mathbf{r}}{r}\right) %
\right]   \notag \\
&+&\frac{q}{16\pi r^{2}}\left[ \overline{\mathbf{E}}\times \mathbf{r}\right] %
\left[ \left( \mathfrak{L}_{\mathfrak{GG}}-\mathfrak{L}_{\mathfrak{FF}%
}\right) \left( \frac{\overline{\mathbf{B}}\cdot \mathbf{r}}{r}\right) +2%
\mathfrak{L}_{\mathfrak{FG}}\left( \frac{\overline{\mathbf{E}}\cdot \mathbf{r%
}}{r}\right) \right]   \notag \\
&+&\frac{q}{8\pi }\left( \mathfrak{L}_{\mathfrak{FF}}+\mathfrak{L}_{%
\mathfrak{GG}}\right) \frac{\left[ \overline{\mathbf{B}}\times \overline{%
\mathbf{E}}\right] }{r^{2}}\,,  \label{4.7}
\end{eqnarray}%
which, in particular, is free of \textquotedblleft string\textquotedblright\
singularities along directions of the external fields, specified by $\xi
,\zeta =\pm 1$. For parallel backgrounds, the cosines $\zeta $, $\xi $
coincide and so do the functions $X\left( \xi \right) $, $Y\left( \zeta
\right) $. In this case, Eq. (\ref{4.7}) reduces to%
\begin{equation}
\mathbf{A}^{\left( 1\right) }\left( \mathbf{r}\right) =-\frac{q}{4\pi }\frac{%
\left[ \mathbf{\hat{z}}\times \mathbf{r}\right] }{r^{2}}\frac{\tilde{g}}{2}%
\zeta \,,  \label{4.12}
\end{equation}%
Note that in this case all the potentials, to which the ansatz (\ref{H})
reduces, obey the Coulomb gauge condition $\mathbf{\nabla }\cdot \mathbf{A}%
^{\left( 1\right) }\left( \mathbf{r}\right) =0$.

\section{Results in QED\label{S5}}

To visualize how the magnetic responses and related effects, valid for any
local nonlinear theory, depend on the constant background, we apply the
former results to a specific theory, whose nonlinearity is provided by the
local approximation of the effective Lagrangian of QED found within
one-fermion-loop calculation by Heisenberg and Euler \cite{Heisenberg} (see
e.g. \cite{BerLifPit}) 
\begin{equation}
\mathfrak{L}=\frac{M^{4}}{8\pi ^{2}}\int_{0}^{\infty }dt\frac{e^{-t}}{t^{3}}%
\left\{ -\left( ta\cot ta\right) \left( tb\coth tb\right) +1-\frac{1}{3}%
\left( a^{2}-b^{2}\right) t^{2}\right\} \,,  \label{eh1}
\end{equation}%
where the integration contour is meant to circumvent the poles on the real
axis of $t$ supplied by $\cot ta$ above the real axis. Here $a$ and $b$ are
dimensionless combinations of the field invariants,%
\begin{eqnarray}
&&a=\frac{e}{M^{2}}\sqrt{-\overline{\mathfrak{F}}+\sqrt{\overline{\mathfrak{F%
}}^{2}+\overline{\mathfrak{G}}^{2}}}\,,  \notag \\
&&b=\frac{e}{M^{2}}\sqrt{\overline{\mathfrak{F}}+\sqrt{\overline{\mathfrak{F}%
}^{2}+\overline{\mathfrak{G}}^{2}}}\,,  \label{eh2}
\end{eqnarray}%
and have the meaning of the electric and magnetic field in the Lorentz frame
where these are parallel, normalized to the characteristic field value $%
M^{2}/e$, where $M$ and $e$ are the electron mass and charge, respectively.
As it is well known, such a frame always exists when $\overline{\mathfrak{G}}%
\neq 0$.

We are primarily interested in strong magnetic-dominated backgrounds, in
which vacuum polarization effects overcome vacuum instability ones. In such
backgrounds, the electric contribution is sufficiently small in comparison
to the magnetic part%
\begin{equation}
\frac{a}{b}\ll 1\,,  \label{ratio}
\end{equation}%
irrespective of whether\ $a$ and $b$ are small or not as compared to the
unity, implying the magnetic dominance $\overline{\mathbf{B}}\gg \overline{%
\mathbf{E}}$ in any reference frame. Such condition is enough to probe
vacuum nonlinear effects\footnote{%
Note that the magnetic responses (\ref{arb21}), (\ref{arb22}) and the
nonlinearly induced currents (\ref{nlin}), (\ref{nlout}), vanish identically
in the pure magnetic background $a/b=0$, since $\mathfrak{G}=0$.}, and
should be applied in final expressions after all coefficients composing the
magnetic responses (derivatives of the effective Lagrangian) have been
calculated (with $a$ and $b$ arbitrary, i.e. not subjected to condition (\ref%
{ratio})). Thus, using general expressions for the derivatives of the
effective Lagrangian, given by Eqs. (52) - (54) in Ref. \cite{AdoGitSha2016}%
, the coefficient $\tilde{g}$ (\ref{gtilded}) takes\ the form%
\begin{eqnarray}
\tilde{g} &=&\frac{M^{4}}{32\pi ^{2}}\frac{b^{2}}{\overline{\mathfrak{F}}%
^{2}+\overline{\mathfrak{G}}^{2}}\int_{0}^{\infty }d\tau \frac{e^{-\tau /b}}{%
\tau ^{3}}\left\{ \mathcal{H}\left( \tau \right) \mathcal{Q}\left( \frac{%
a\tau }{b}\right) \left( -4\overline{\mathfrak{G}}+2\overline{\mathfrak{F}}%
\kappa \left( \frac{a}{b}-\frac{b}{a}\right) \right) \right.  \notag \\
&+&\tau \coth \tau \left[ \left( 2\overline{\mathfrak{G}}\gamma _{+}-2%
\overline{\mathfrak{F}}\kappa \frac{b}{a}\gamma _{-}\right) \mathcal{Q}%
\left( \frac{a\tau }{b}\right) \right.  \notag \\
&+&\left. \left( \overline{\mathfrak{G}}\left( 1-\frac{b^{2}}{a^{2}}\right)
+2\overline{\mathfrak{F}}\kappa \frac{b}{a}\right) \mathcal{\tilde{Q}}\left( 
\frac{a\tau }{b}\right) \right]  \notag \\
&+&\left( \frac{a\tau }{b}\cot \frac{a\tau }{b}\right) \left[ 2\left( 
\overline{\mathfrak{G}}\gamma _{-}+\overline{\mathfrak{F}}\kappa \gamma _{+}%
\frac{a}{b}\right) \mathcal{H}\left( \tau \right) \right.  \notag \\
&+&\left. \left. \left( \overline{\mathfrak{G}}\left( 1-\frac{a^{2}}{b^{2}}%
\right) -2\overline{\mathfrak{F}}\kappa \frac{a}{b}\right) \mathcal{\tilde{H}%
}\left( \tau \right) \right] \right\} \,,  \label{eh25}
\end{eqnarray}%
where $\mathcal{H}\left( \tau \right) $, $\mathcal{Q}\left( \tau \right) $, $%
\mathcal{\tilde{H}}\left( \tau \right) $, $\mathcal{\tilde{Q}}\left( \tau
\right) $ are auxiliary functions%
\begin{eqnarray}
\mathcal{H}\left( \tau \right) &=&\tau \coth \tau -\frac{\tau ^{2}}{\sinh
^{2}\tau }\,,\ \ \mathcal{Q}\left( \tau \right) =\tau \cot \tau -\frac{\tau
^{2}}{\sin ^{2}\tau }\,,  \notag \\
\mathcal{\tilde{H}}\left( \tau \right) &=&\frac{2\tau ^{2}}{\sinh ^{2}\tau }%
\left( \tau \coth \tau -1\right) \,,\ \ \mathcal{\tilde{Q}}\left( \tau
\right) =\frac{2\tau ^{2}}{\sin ^{2}\tau }\left( \tau \cot \tau -1\right) \,;
\label{eh27}
\end{eqnarray}%
$\kappa =\mathrm{sgn}\left( \overline{\mathfrak{G}}\right) $ and $\gamma
_{\pm }=1\pm 2\overline{\mathfrak{F}}\mathfrak{/}\sqrt{\overline{\mathfrak{F}%
}^{2}+\overline{\mathfrak{G}}^{2}}$ are constants. Eq. (\ref{eh25}) together
with Eqs. (52) - (54) from Ref. \cite{AdoGitSha2016}, provide integral
representations for all the necessary coefficients within the
Euler-Heisenberg nonlinear electrodynamics to be substituted into Eqs. (\ref%
{arb21})-(\ref{arb22R}) for specializing the magnetic responses.

Let us study the strong magnetic-dominated case, specified by the condition (%
\ref{ratio}). To this end, we rewrite the coefficient (\ref{gtilded}) as $%
\tilde{g}=-\overline{\mathfrak{G}}\mathfrak{L}_{-}+2\overline{\mathfrak{F}}%
\mathfrak{L}_{\mathfrak{FG}}$\thinspace , $\mathfrak{L}_{-}=\mathfrak{L}_{%
\mathfrak{FF}}-\mathfrak{L}_{\mathfrak{GG}}$, and expand trigonometric
functions within $\mathfrak{L}_{-}$, $\mathfrak{L}_{\mathfrak{FG}}$ in power
series of $a/b$ (avoiding the poles at the real axis in Eq. (\ref{eh25}),
thereby). The first coefficient $\overline{\mathfrak{G}}\mathfrak{L}_{-}$
reads%
\begin{eqnarray}
&&\overline{\mathfrak{G}}\mathfrak{L}_{-}=\left( \frac{a}{b}\right) \left( 
\overline{\mathfrak{G}}\mathfrak{L}_{-}\right) ^{\left( 1\right) }+O\left(
\left( a/b\right) ^{3}\right) \,,  \notag \\
&&\left( \overline{\mathfrak{G}}\mathfrak{L}_{-}\right) ^{\left( 1\right)
}=2\kappa \left( \frac{\alpha }{2\pi }\right) \int_{0}^{\infty }d\tau \frac{%
e^{-\tau /b}}{\tau }\left[ \left( \frac{1}{\tau }-\frac{\tau }{3}\right)
\coth \tau -\tau \frac{\coth \tau }{\sinh ^{2}\tau }\right] \,,  \label{ehm3}
\end{eqnarray}%
while the second $\overline{\mathfrak{F}}\mathfrak{L}_{\mathfrak{FG}}$ takes
the form,%
\begin{eqnarray}
&&\overline{\mathfrak{F}}\mathfrak{L}_{\mathfrak{FG}}=\left( \frac{a}{b}%
\right) \left( \overline{\mathfrak{F}}\mathfrak{L}_{\mathfrak{FG}}\right)
^{\left( 1\right) }+O\left( \left( a/b\right) ^{3}\right) \,,  \notag \\
&&\left( \overline{\mathfrak{F}}\mathfrak{L}_{\mathfrak{FG}}\right) ^{\left(
1\right) }=\kappa \left( \frac{\alpha }{2\pi }\right) \int_{0}^{\infty
}d\tau \frac{e^{-\tau /b}}{\tau }\left[ \left( \frac{3}{\tau }-\frac{2\tau }{%
3}\right) \coth \tau -\left( 1+\frac{2\tau ^{2}}{3}\right) \frac{1}{\sinh
^{2}\tau }-2\tau \frac{\coth \tau }{\sinh ^{2}\tau }\right] \,.  \label{ehm4}
\end{eqnarray}%
As a result, the leading-order contribution for $\tilde{g}$ is expressed as%
\begin{eqnarray}
&&\tilde{g}=\left( \frac{a}{b}\right) \tilde{g}^{\left( 1\right) }+O\left(
\left( a/b\right) ^{3}\right) \,,  \notag \\
&&\tilde{g}^{\left( 1\right) }=2\kappa \left( \frac{\alpha }{2\pi }\right)
\int_{0}^{\infty }d\tau \frac{e^{-\tau /b}}{\tau }\left[ \left( \frac{2}{%
\tau }-\frac{\tau }{3}\right) \coth \tau -\left( 1+\frac{2\tau ^{2}}{3}%
\right) \frac{1}{\sinh ^{2}\tau }-\tau \frac{\coth \tau }{\sinh ^{2}\tau }%
\right] \,.  \label{ehm5}
\end{eqnarray}

To estimate the asymptotic behavior of $\tilde{g}$ (\ref{ehm5}) in the
infinite magnetic field limit $b\rightarrow \infty $, it is convenient to\
express all integrals above in terms of the Hurwitz Zeta function\ $\zeta
\left( z,a\right) $, DiGamma function $\psi \left( z\right) =\Gamma ^{\prime
}\left( z\right) /\Gamma \left( z\right) $ and related functions \cite{DLMF}%
. Using Zeta-function regularization techniques (see e. g., \cite%
{Dittrich,Elizalde,Klaus}), the expansion coefficient is expressed as follows%
\begin{eqnarray}
\tilde{g}^{\left( 1\right) } &=&2\kappa \left( \frac{\alpha }{2\pi }\right)
\left\{ \frac{b}{3}+\frac{1}{6}-12\zeta ^{\prime }\left( -1,\frac{1}{2b}%
\right) +\psi \left( \frac{1}{2b}\right) +\frac{1}{2b^{2}}\left[ \psi \left( 
\frac{1}{2b}\right) +\frac{1}{2}\right] \right.  \notag \\
&+&\frac{1}{b}\left. \left[ 2\log 2b-\log 2\pi +\frac{1}{2}+\frac{1}{3}\psi
^{(1)}\left( \frac{1}{2b}\right) +2\log \Gamma \left( \frac{1}{2b}\right) %
\right] \right\} \,,  \label{ehm7}
\end{eqnarray}%
where $\zeta ^{\prime }\left( z,a\right) $ is the derivative of the Hurwitz
Zeta function with respect to $z$, $\psi ^{\left( j\right) }\left( z\right) $
is the $j$-th derivative of the DiGamma function and $\gamma \approx 0.577$
is the Euler constant \cite{DLMF}. Using asymptotic representations of
special functions (\ref{apb8}), (\ref{apb9}), discussed in Appendix \ref%
{AppeB}, the above coefficient behaves as%
\begin{eqnarray}
&&\tilde{g}^{\left( 1\right) }\sim 2\kappa \left( \frac{\alpha }{2\pi }%
\right) \left( -\frac{b}{3}+\mathcal{K}_{\tilde{g}}^{\left( 1\right)
}\right) \,,\ \ b\rightarrow \infty \,,  \notag \\
&&\mathcal{K}_{\tilde{g}}^{\left( 1\right) }=\frac{1}{6}-\gamma -12\zeta
^{\prime }\left( -1\right) \,,  \label{em8}
\end{eqnarray}%
in the large-field limit. Therefore it is easily seen that the leading-order
contribution to $\tilde{g}$%
\begin{equation}
\tilde{g}\sim 2\kappa \left( \frac{\alpha }{2\pi }\right) \left( -\frac{a}{3}%
+\frac{a}{b}\mathcal{K}_{\tilde{g}}^{\left( 1\right) }\right) \,,\ \
b\rightarrow \infty \,.  \label{ehm9}
\end{equation}%
is proportional to the electric part $a$. The pseudoscalar quality to $%
\tilde{g}$ is imparted by the factor $\kappa =\mathrm{sgn\,}\overline{%
\mathfrak{G}}$, because $\overline{\mathfrak{G}}$ changes its sign under
spacial reflection.

\section{Conclusions\label{Conc}}

Within a nonlinear local electrodynamics (\ref{L}), we have obtained
magnetic fields created by a static electric charge $q$ placed in a
background of arbitrarily strong constant electric, $\overline{\mathbf{E}}$,
and magnetic, $\overline{\mathbf{B}}$, fields by solving (the second pair
of) the Maxwell equations (\ref{s1.51}) linearized near the background and
treated in the approximation of small nonlinearity. All our formulas contain
coefficients that are derivatives of the nonlinear part of the Lagrangian (%
\ref{L}), where the background values of the fields are meant to be
substituted after the derivatives have been calculated. These coefficients
are related to dielectric permeability and magnetic permittivity of the
equivalent \textquotedblleft medium\textquotedblright\ formed by the
background fields in the vacuum \cite{ChavShab}.

Before considering the necessary magnetic fields we establish the character
of their source, which comprises of the currents induced in the equivalent
\textquotedblleft medium\textquotedblright\ by the static charge. The result
for the current inside and outside of the charge is given by Eqs. (\ref{nlin}%
), (\ref{nlout}). The flow of this current (\ref{par}) for the special case
of parallel background fields is shown in Fig. \ref{Fig1}. There is no
induced current inside the charge in this special case.

The magnetic response $\mathbf{B}^{\left( 1\right) }\left( \mathbf{r}\right) 
$ to an introduced small extended electric charge (\ref{const charge})
homogeneously distributed over a sphere of the radius $R$ is given by Eq. (%
\ref{arb21}) inside the charged sphere and by Eqs. (\ref{arb220}), (\ref%
{arb22R}) outside it. The case of the pointlike charge (\ref{pointlike}) is
covered by the $R\rightarrow 0$ limit, Eq. (\ref{arb220}). In the simplified
case of parallel background fields $\overline{\mathbf{B}}\parallel \overline{%
\mathbf{E}}$, the magnetic response is given by Eqs. (\ref{arb23.1}), (\ref%
{arb23.2}) inside and outside the extended charge, respectively, and by Eq. (%
\ref{ansa20}) for the point charge. The pattern of magnetic lines of force
is presented in Fig.\ref{Fig2}. To complete the study, vector potentials
associated with magnetic responses generated by a pointlike Coulomb source
were caculated in Sec. \ref{S4}.

Adjusting the present results with the realistic situation where the
nonlinearity of the Maxwell equations is owing to nonlinearity stemming from
the quantum interaction between electromagnetic fields inherent to QED, we
give integral representations for all the nonlinearity coefficients, in
terms of which our results for the fields are expressed, as they follow from
appropriate differentiations with respect to the field invariants of the
effective Lagrangian of QED in its local approximation taken as the
Euler-Heisenberg (one-loop) effective Lagrangian. We consider the asymptotic
regime when the magnetic background dominates over the electric one. We
found that in that regime the above integrals are conveniently expressed in
terms of the Hurwitz Zeta function. The resulting formula for $\tilde{g}$ (%
\ref{gtilded}), a common coefficient in magnetic responses, linearly induced
current densities and vector potentials, is linear in the background
electric field.

\section*{Acknowledgements}

The work is supported by RFBR under Project 17-02-00317, and by the Tomsk State University Competitiveness Improvement Program. T.C.A. also thanks the Advanced Talents Development Program of the Hebei University, project no. 801260201271, for the partial support. D.M.G. is also supported by the Grant no. 2016/03319-6, Funda\c{c}\~{a}o de Amparo \`{a} Pesquisa do
Estado de S\~{a}o Paulo (FAPESP), and permanently by Conselho Nacional de
Desenvolvimento Cient\'{\i}fico e Tecnol\'{o}gico (CNPq), Brazil.

\begin{appendices}

\section{Projection operator\label{ApA}}

In this Appendix we present further details concerning the derivation of the
first-order linear magnetic responses $\mathbf{B}^{\left( 1\right) }\left( 
\mathbf{r}\right) $ to the static charge in consideration, given by Eqs. (%
\ref{arb21}) and (\ref{arb22}). Referring to Eqs. (\ref{nnew13b}) and (\ref%
{in2}),\ the integral in (\ref{sc2.5b}) can be expressed in terms of
auxiliary integrals $\mathcal{I}^{j}\left( \mathbf{r}\right) $,%
\begin{eqnarray}
\mathcal{I}^{j}\left( \mathbf{r}\right) &=&w\left( r\right) x^{j}\,,\ \
w\left( r\right) =\frac{1}{r^{2}}\int d\mathbf{y}\frac{\Phi \left( y\right)
\left( \mathbf{y\cdot r}\right) }{\left\vert \mathbf{r-y}\right\vert }\,, 
\notag \\
\Phi \left( y\right) &=&\frac{\theta \left( R-y\right) }{R^{3}}+\frac{\theta
\left( y-R\right) }{y^{3}}\,,\ \ r=\left\vert \mathbf{r}\right\vert \,,\ \
y=\left\vert \mathbf{y}\right\vert \,,  \label{apa0}
\end{eqnarray}%
in which the function $\Phi \left( r\right) $, stemming from the regularized
Coulomb field $\mathbf{E}^{(0)}\left( \mathbf{r}\right) $ (\ref{in2}),
provides magnetic responses over all points of the space. From the latter,
the integral in the rhs. of Eq. (\ref{sc2.5b}) is expressed as%
\begin{eqnarray}
&&\int d\mathbf{y}\frac{\mathfrak{H}^{(0)j}\left( \mathbf{y}\right) }{%
\left\vert \mathbf{r-y}\right\vert }=-\frac{q}{4\pi }\mathfrak{L}_{\mathfrak{%
G}}\mathcal{I}^{j}\left( \mathbf{r}\right) -\frac{q}{4\pi }\mathfrak{T}%
_{B}^{jk}\mathcal{I}^{k}\left( \mathbf{r}\right) \,,  \notag \\
&&\mathfrak{T}_{B}^{jk}=\left( \mathfrak{L}_{\mathfrak{FF}}\overline{B}^{j}-%
\mathfrak{L}_{\mathfrak{FG}}\overline{E}^{j}\right) \overline{E}^{k}+\left( 
\mathfrak{L}_{\mathfrak{FG}}\overline{B}^{j}-\mathfrak{L}_{\mathfrak{GG}}%
\overline{E}^{j}\right) \overline{B}^{k}\,.  \label{arb17}
\end{eqnarray}%
Next, one can evaluate the scalar integral $w\left( r\right) $ by selecting
a system of reference in which $\mathbf{r}$ is aligned along the $z$%
-direction, such that $\mathbf{y\cdot r}=yr\cos \theta $ and it is reduced
to a simple radial integral%
\begin{eqnarray*}
w\left( r\right) &=&\frac{2\pi }{r}\int_{0}^{\infty }dyy^{3}\Phi \left(
y\right) \int_{-1}^{1}d\left( \cos \theta \right) \frac{\cos \theta }{\sqrt{%
r^{2}+y^{2}-2ry\cos \theta }} \\
&=&\frac{2\pi }{3r^{3}}\int_{0}^{\infty }dyy\Phi \left( y\right) \left\{
\left( r^{2}-ry+y^{2}\right) \left( r+y\right) -\left( r^{2}+ry+y^{2}\right)
\left\vert r-y\right\vert \right\} \,.
\end{eqnarray*}%
Computing the remaining integrals above, the auxiliary integrals (\ref{apa0}%
) acquires the final form%
\begin{eqnarray}
\mathcal{I}^{j}\left( \mathbf{r}\right) &=&2\pi \varrho \left( r\right)
x^{j}\,,\ \ \varrho \left( r\right) =\varrho _{\mathrm{in}}\left( r\right)
\theta \left( R-r\right) +\varrho _{\mathrm{out}}\left( r\right) \theta
\left( r-R\right) \,,  \notag \\
\varrho _{\mathrm{in}}\left( r\right) &=&\frac{1}{R}\left( 1-\frac{r^{2}}{%
5R^{2}}\right) \,,\ \ \varrho _{\mathrm{out}}\left( r\right) =\frac{1}{r}%
\left( 1-\frac{R^{2}}{5r^{2}}\right) \,.  \label{apa2}
\end{eqnarray}

From the definition of $\mathfrak{T}_{B}^{jk}$ (\ref{arb17}), one may use
the following set of identities%
\begin{eqnarray}
&&\mathfrak{T}_{B}^{ik}x^{k}=\left[ \mathfrak{L}_{\mathfrak{FF}}\left( 
\overline{\mathbf{E}}\cdot \mathbf{r}\right) +\mathfrak{L}_{\mathfrak{FG}%
}\left( \mathbf{\overline{B}}\cdot \mathbf{r}\right) \right] \overline{B}%
^{i}-\left[ \mathfrak{L}_{\mathfrak{FG}}\left( \overline{\mathbf{E}}\cdot 
\mathbf{r}\right) +\mathfrak{L}_{\mathfrak{GG}}\left( \mathbf{\overline{B}}%
\cdot \mathbf{r}\right) \right] \overline{E}^{i}\,,  \notag \\
&&\mathfrak{T}_{B}^{ki}x^{k}=\left[ \mathfrak{L}_{\mathfrak{FF}}\left( 
\mathbf{\overline{B}}\cdot \mathbf{r}\right) -\mathfrak{L}_{\mathfrak{FG}%
}\left( \overline{\mathbf{E}}\cdot \mathbf{r}\right) \right] \overline{E}%
^{i}+\left[ \mathfrak{L}_{\mathfrak{FG}}\left( \mathbf{\overline{B}}\cdot 
\mathbf{r}\right) -\mathfrak{L}_{\mathfrak{GG}}\left( \overline{\mathbf{E}}%
\cdot \mathbf{r}\right) \right] \overline{B}^{i}\,,  \notag \\
&&\mathfrak{T}_{B}^{jk}x^{i}\delta ^{jk}=\mathfrak{T}_{B}^{jj}x^{i}=\tilde{g}%
x^{i}\,,\ \ \mathfrak{T}_{B}^{jk}\left( \delta ^{ij}x^{k}+x^{j}\delta
^{ik}+x^{i}\delta ^{jk}\right) =U_{E}\overline{E}^{i}+U_{B}\overline{B}^{i}+%
\tilde{g}x^{i}\,,  \notag \\
&&U_{E}=\left( \mathfrak{L}_{\mathfrak{FF}}-\mathfrak{L}_{\mathfrak{GG}%
}\right) \left( \mathbf{\overline{B}}\cdot \mathbf{r}\right) -2\mathfrak{L}_{%
\mathfrak{FG}}\left( \overline{\mathbf{E}}\cdot \mathbf{r}\right) \,,  \notag
\\
&&U_{B}=\left( \mathfrak{L}_{\mathfrak{FF}}-\mathfrak{L}_{\mathfrak{GG}%
}\right) \left( \overline{\mathbf{E}}\cdot \mathbf{r}\right) +2\mathfrak{L}_{%
\mathfrak{FG}}\left( \mathbf{\overline{B}}\cdot \mathbf{r}\right) \,,
\label{arb18}
\end{eqnarray}%
to learn that the action of partial derivatives on Eq. (\ref{arb17}) takes
the final form%
\begin{eqnarray}
\frac{\partial _{i}\partial _{j}}{4\pi }\int d\mathbf{y}\frac{\mathfrak{H}%
^{\left( 0\right) j}\left( \mathbf{y}\right) }{\left\vert \mathbf{r-y}%
\right\vert } &=&\left( \frac{\mathfrak{L}_{\mathfrak{G}}}{3}-\frac{qU_{E}}{%
8\pi }\frac{\varrho ^{\prime }\left( r\right) }{r}\right) \overline{E}%
^{i}-\left( \frac{\mathfrak{L}_{\mathfrak{F}}}{3}+\frac{qU_{B}}{8\pi }\frac{%
\varrho ^{\prime }\left( r\right) }{r}\right) \overline{B}^{i}  \notag \\
&-&\frac{q}{8\pi }\left[ \frac{\tilde{g}\varrho ^{\prime }\left( r\right) }{r%
}+\mathfrak{L}_{\mathfrak{G}}\left( \frac{4\varrho ^{\prime }\left( r\right) 
}{r}+\varrho ^{\prime \prime }\left( r\right) \right) +\left( \varrho
^{\prime \prime }\left( r\right) -\frac{\varrho ^{\prime }\left( r\right) }{r%
}\right) \frac{x^{j}\mathfrak{T}_{B}^{jk}x^{k}}{r^{2}}\right] x^{i}\,.
\label{arb19}
\end{eqnarray}

It should be noted that the inner $\varrho _{\mathrm{in}}\left( r\right) $
and the outer $\varrho _{\mathrm{out}}\left( r\right) $ components of $%
\varrho \left( r\right) $, defined in Eq. (\ref{apa2}), are continuous at $%
r=R$ (as well as its first and second derivatives). For these reasons,
coefficients proportional to Dirac delta functions, stemming from the
differentiation of Heaviside step functions, vanishes everywhere, including
at $r=R$. Accordingly, derivatives of $\varrho \left( r\right) $ can be
treated as $\varrho ^{\prime }\left( r\right) =\varrho _{\mathrm{in}%
}^{\prime }\left( r\right) \theta \left( R-r\right) +\varrho _{\mathrm{out}%
}^{\prime }\left( r\right) \theta \left( r-R\right) $ and $\varrho ^{\prime
\prime }\left( r\right) =\varrho _{\mathrm{in}}^{\prime \prime }\left(
r\right) \theta \left( R-r\right) +\varrho _{\mathrm{out}}^{\prime \prime
}\left( r\right) \theta \left( r-R\right) $.

\section{Expansion coefficients\label{AppeB}}

In this Appendix we present exact expressions for the expansion coefficients
of the derivatives of the Heisenberg-Euler effective Lagrangian (\ref{eh1})
in the strong magnetic-dominated case, discussed in Sec. \ref{S5}, in terms
of the Hurwitz Zeta and related functions. Moreover, we supplement some of
our previous results concerning the electric response to the charge
distribution (\ref{const charge}) by the background in consideration,
namely, the nonlinearly induced total charge inside the distribution $Q$ and
the coefficient $\tilde{b}$, both given by Eqs. (32) and (33) in Ref. \cite%
{AdoGitSha2016}, respectively. Using formulae below, we write the leading
and next-to-leading expansion coefficients for these quantities.

Starting with parity-even coefficients $\mathcal{S}_{\mathrm{e}}=\left\{ 
\mathfrak{L}_{\mathfrak{F}},\mathfrak{L}_{\mathfrak{FF}},\mathfrak{L}_{%
\mathfrak{GG}}\right\} $, which admits expansions of the form%
\begin{equation}
\mathcal{S}_{\mathrm{e}}=\mathcal{S}_{\mathrm{e}}^{\left( 0\right) }+\left( 
\frac{a}{b}\right) ^{2}\mathcal{S}_{\mathrm{e}}^{\left( 2\right) }+O\left(
\left( \frac{a}{b}\right) ^{4}\right) \,,  \label{PEC}
\end{equation}%
the corresponding leading and next-to-leading contributions have the form:%
\begin{eqnarray}
\mathfrak{L}_{\mathfrak{F}}^{\left( 0\right) } &=&\frac{\alpha }{2\pi }%
\left\{ -\frac{1}{2b^{2}}-\frac{1}{3}+\frac{2}{3}\log 2+\frac{2}{3}\log b+%
\frac{1}{b}\log \left( \frac{\pi }{b}\right) \right.  \notag \\
&-&\left. \frac{2}{b}\log \Gamma \left( \frac{1}{2b}\right) +8\zeta ^{\prime
}\left( -1,\frac{1}{2b}\right) \right\}  \notag \\
\mathfrak{L}_{\mathfrak{F}}^{\left( 2\right) } &=&\frac{\alpha }{2\pi }%
\left\{ \frac{b}{3}+\frac{1}{2b^{2}}+\frac{1}{b}\log \left( \frac{b}{\pi }%
\right) +\frac{2}{b}\log \Gamma \left( \frac{1}{2b}\right) +\frac{2}{3}\psi
\left( \frac{1}{2b}\right) \right.  \notag \\
&+&\left. \frac{1}{6b}\psi ^{\left( 1\right) }\left( \frac{1}{2b}\right)
-8\zeta ^{\prime }\left( -1,\frac{1}{2b}\right) \right\} \,,  \label{apb1}
\end{eqnarray}%
and%
\begin{eqnarray}
\left( \overline{\mathfrak{F}}+\sqrt{\overline{\mathfrak{F}}^{2}+\overline{%
\mathfrak{G}}^{2}}\right) \mathfrak{L}_{\mathfrak{FF}}^{\left( 0\right) } &=&%
\frac{\alpha }{2\pi }\left\{ \frac{2}{3}-\frac{1}{b^{2}}+\frac{1}{b}+\frac{1%
}{b}\log \frac{4b}{\pi }-\frac{2}{b}\log \Gamma \left( \frac{1}{2b}\right) +%
\frac{1}{b^{2}}\psi \left( \frac{1}{2b}\right) \right\}  \notag \\
\left( \overline{\mathfrak{F}}+\sqrt{\overline{\mathfrak{F}}^{2}+\overline{%
\mathfrak{G}}^{2}}\right) \mathfrak{L}_{\mathfrak{FF}}^{\left( 2\right) } &=&%
\frac{\alpha }{2\pi }\left\{ -\frac{2}{3}-b-\frac{8}{3}\psi \left( \frac{1}{%
2b}\right) +32\zeta ^{\prime }\left( -1,\frac{1}{2b}\right) \right.  \notag
\\
&+&\frac{1}{b}\left[ -2-\frac{7}{6}\psi ^{\left( 1\right) }\left( \frac{1}{2b%
}\right) -6\log 2b+2\log 2\pi -4\log \Gamma \left( \frac{1}{2b}\right) %
\right]  \notag \\
&+&\frac{1}{b^{2}}\left. \left[ \frac{1}{6}\zeta \left( 3,\frac{1}{2b}%
\right) -2\psi \left( \frac{1}{2b}\right) \right] \right\}  \notag \\
\left( \overline{\mathfrak{F}}+\sqrt{\overline{\mathfrak{F}}^{2}+\overline{%
\mathfrak{G}}^{2}}\right) \mathfrak{L}_{\mathfrak{GG}}^{\left( 0\right) } &=&%
\frac{\alpha }{2\pi }\left\{ -\frac{2b}{3}-\frac{1}{3}-\frac{1}{2b^{2}}%
\right.  \notag \\
&+&\left. \frac{1}{b}\left[ \log \left( \frac{\pi }{b}\right) -2\log \Gamma
\left( \frac{1}{2b}\right) \right] +8\zeta ^{\prime }\left( -1,\frac{1}{2b}%
\right) -\frac{2}{3}\psi \left( \frac{1}{2b}\right) \right\}  \notag \\
\left( \overline{\mathfrak{F}}+\sqrt{\overline{\mathfrak{F}}^{2}+\overline{%
\mathfrak{G}}^{2}}\right) \mathfrak{L}_{\mathfrak{GG}}^{\left( 2\right) } &=&%
\frac{\alpha }{2\pi }\left\{ -\frac{8b^{3}}{15}+\frac{5b}{3}+\frac{2}{3}%
-40\zeta ^{\prime }\left( -1,\frac{1}{2b}\right) +\frac{2}{15}\zeta \left( 3,%
\frac{1}{2b}\right) \right.  \notag \\
&+&\frac{10}{3}\psi \left( \frac{1}{2b}\right) +\frac{1}{b^{2}}\left[ \frac{3%
}{2}+\psi \left( \frac{1}{2b}\right) \right]  \notag \\
&+&\left. \frac{1}{b}\left[ 1+6\log 2b-4\log 2\pi +8\log \Gamma \left( \frac{%
1}{2b}\right) +\frac{5}{6}\psi ^{\left( 1\right) }\left( \frac{1}{2b}\right) %
\right] \right\} \,.  \label{apb2}
\end{eqnarray}%
The parity-even combination $\overline{\mathfrak{G}}\mathfrak{L}_{\mathfrak{%
FG}}$ has a trivial zero-order term%
\begin{eqnarray}
\overline{\mathfrak{G}}\mathfrak{L}_{\mathfrak{FG}} &=&\left( \frac{a}{b}%
\right) ^{2}\left( \overline{\mathfrak{G}}\mathfrak{L}_{\mathfrak{FG}%
}\right) ^{\left( 2\right) }+O\left( \left( \frac{a}{b}\right) ^{4}\right)
\,,  \notag \\
\left( \overline{\mathfrak{G}}\mathfrak{L}_{\mathfrak{FG}}\right) ^{\left(
2\right) } &=&\frac{\alpha }{2\pi }\left\{ \frac{2}{3}+\frac{2b}{3}-16\zeta
^{\prime }\left( -1,\frac{1}{2b}\right) +\frac{4}{3}\psi \left( \frac{1}{2b}%
\right) +\frac{1}{b^{2}}\psi \left( \frac{1}{2b}\right) \right.  \notag \\
&+&\left. \frac{1}{b}\left[ 1+\frac{1}{3}\psi ^{\left( 1\right) }\left( 
\frac{1}{2b}\right) +3\log 2b-\log 2\pi +2\log \Gamma \left( \frac{1}{2b}%
\right) \right] \right\} \,.  \label{apb4}
\end{eqnarray}%
because it must vanish identically if the electric background is zero. The
proper-time representations to all these coefficients are given by Eqs.
(57), (58) in Ref. \cite{AdoGitSha2016}. For the sake of completess, we also
include the parity-odd coefficient $\mathfrak{L}_{\mathfrak{G}}$%
\begin{eqnarray}
\mathfrak{L}_{\mathfrak{G}} &=&-\frac{M^{4}}{16\pi ^{2}}\frac{\kappa b^{2}}{%
\sqrt{\overline{\mathfrak{F}}^{2}+\overline{\mathfrak{G}}^{2}}}\left( \frac{a%
}{b}\right) \int_{0}^{\infty }d\tau \frac{e^{-\tau /b}}{\tau ^{3}}\left\{
\left( \frac{b}{a}\right) ^{2}\left( \tau \coth \tau \right) \mathcal{Q}%
\left( \frac{a\tau }{b}\right) \right.  \notag \\
&+&\left. \left( \frac{a\tau }{b}\cot \frac{a\tau }{b}\right) \mathcal{H}%
\left( \tau \right) \right\} \,,  \label{LG}
\end{eqnarray}%
which it is expanded as%
\begin{eqnarray}
\mathfrak{L}_{\mathfrak{G}} &=&\left( \frac{a}{b}\right) \mathfrak{L}_{%
\mathfrak{G}}^{\left( 1\right) }+O\left( \left( \frac{a}{b}\right)
^{3}\right) \,,  \notag \\
\mathfrak{L}_{\mathfrak{G}}^{\left( 1\right) } &=&-\kappa \left( \frac{%
\alpha }{2\pi }\right) \int_{0}^{\infty }d\tau \frac{e^{-\tau /b}}{\tau }%
\left[ \left( \frac{1}{\tau }-\frac{2}{3}\tau \right) \coth \tau -\frac{1}{%
\sinh ^{2}\tau }\right]  \notag \\
&=&-\kappa \left( \frac{\alpha }{2\pi }\right) \left\{ \frac{2b}{3}+\frac{1}{%
3}+\frac{2}{3}\psi \left( \frac{1}{2b}\right) -8\zeta ^{\prime }\left( -1,%
\frac{1}{2b}\right) \right.  \notag \\
&+&\left. \frac{1}{b}\left[ \log \frac{b}{\pi }+2\log \Gamma \left( \frac{1}{%
2b}\right) \right] +\frac{1}{2b^{2}}\right\} \,,  \label{apb3}
\end{eqnarray}

With the help of these coefficients, the nonlinearly induced electric charge 
$Q$ and the coefficient $\tilde{b}$, corresponding to the electric response
of the background to the charge (\ref{const charge}),%
\begin{equation}
Q=q\left( \mathfrak{L}_{\mathfrak{F}}+\frac{\tilde{b}}{3}\right) \,,\ \ 
\tilde{b}=-\left( \mathfrak{L}_{\mathfrak{FF}}\mathbf{E}^{2}+\mathfrak{L}_{%
\mathfrak{GG}}\mathbf{B}^{2}-2\overline{\mathfrak{G}}\mathfrak{L}_{\mathfrak{%
FG}}\right) \,,  \label{apb5}
\end{equation}%
are expanded as follows:%
\begin{eqnarray}
\tilde{b} &=&\tilde{b}^{\left( 0\right) }+\left( \frac{a}{b}\right) ^{2}%
\tilde{b}^{\left( 2\right) }+O\left( \left( \frac{a}{b}\right) ^{4}\right)
\,,  \notag \\
\tilde{b}^{\left( 0\right) } &=&\overline{\mathfrak{F}}\mathfrak{L}%
_{-}^{\left( 0\right) }-\sqrt{\overline{\mathfrak{F}}^{2}+\overline{%
\mathfrak{G}}^{2}}\mathfrak{L}_{+}^{\left( 0\right) }\,,\ \ \tilde{b}%
^{\left( 2\right) }=\overline{\mathfrak{F}}\mathfrak{L}_{-}^{\left( 2\right)
}-\sqrt{\overline{\mathfrak{F}}^{2}+\overline{\mathfrak{G}}^{2}}\mathfrak{L}%
_{+}^{\left( 2\right) }+2\overline{\mathfrak{G}}\mathfrak{L}_{\mathfrak{FG}%
}^{\left( 2\right) }\,,  \notag \\
\tilde{b}^{\left( 0\right) } &=&\frac{\alpha }{2\pi }\int_{0}^{\infty }d\tau 
\frac{e^{-\tau /b}}{\tau }\left[ \left( \frac{1}{\tau }-\frac{2\tau }{3}%
\right) \coth \tau -\frac{1}{\sinh ^{2}\tau }\right]  \notag \\
&=&\frac{\alpha }{2\pi }\left\{ \frac{2b}{3}+\frac{1}{3}+\frac{2}{3}\psi
\left( \frac{1}{2b}\right) -8\zeta ^{\prime }\left( -1,\frac{1}{2b}\right)
\right.  \notag \\
&+&\left. \frac{1}{b}\left[ \log \left( \frac{b}{\pi }\right) +2\log \Gamma
\left( \frac{1}{2b}\right) \right] +\frac{1}{2b^{2}}\right\}  \notag \\
\tilde{b}^{\left( 2\right) } &=&\frac{\alpha }{2\pi }\int_{0}^{\infty }d\tau 
\frac{e^{-\tau /b}}{\tau }\left[ \left( \frac{\tau }{3}-\frac{4\tau ^{3}}{15}%
-\frac{1}{\tau }\right) \coth \tau +\left( 1+\frac{\tau ^{2}}{3}\right) 
\frac{1}{\sinh ^{2}\tau }\right]  \notag \\
&=&\frac{\alpha }{2\pi }\left\{ \frac{8b^{3}}{15}-\frac{b}{3}-\frac{2}{15}%
\zeta \left( 3,\frac{1}{2b}\right) -\frac{2}{3}\psi \left( \frac{1}{2b}%
\right) +8\zeta ^{\prime }\left( -1,\frac{1}{2b}\right) \right.  \notag \\
&-&\left. \frac{1}{b}\left[ \log \left( \frac{b}{\pi }\right) +\frac{1}{6}%
\psi ^{(1)}\left( \frac{1}{2b}\right) +2\log \Gamma \left( \frac{1}{2b}%
\right) \right] -\frac{1}{2b^{2}}\right\} \,,  \label{apb6}
\end{eqnarray}%
where $\mathfrak{L}_{\pm }=\mathfrak{L}_{\mathfrak{FF}}\pm \mathfrak{L}_{%
\mathfrak{GG}}$ and%
\begin{eqnarray}
Q &=&Q^{\left( 0\right) }+\left( \frac{a}{b}\right) ^{2}Q^{\left( 2\right)
}+O\left( \left( \frac{a}{b}\right) ^{4}\right) \,,\ \ Q^{\left( i\right)
}=q\left( \mathfrak{L}_{\mathfrak{F}}^{\left( i\right) }+\frac{\tilde{b}%
^{\left( i\right) }}{3}\right) \,,  \notag \\
Q^{\left( 0\right) } &=&q\left( \frac{\alpha }{3\pi }\right)
\int_{0}^{\infty }d\tau \frac{e^{-\tau /b}}{\tau }\left\{ 1-\left( \frac{1}{%
\tau }+\frac{\tau }{3}\right) \coth \tau +\frac{1}{\sinh ^{2}\tau }\right\}
\,,  \notag \\
&=&q\left( \frac{\alpha }{3\pi }\right) \left\{ \frac{b}{3}+\log b-\frac{1}{3%
}+\log 2+\frac{1}{3}\psi \left( \frac{1}{2b}\right) +8\zeta ^{\prime }\left(
-1,\frac{1}{2b}\right) \right.  \notag \\
&+&\left. \frac{1}{b}\left[ \log \left( \frac{\pi }{b}\right) -2\log \Gamma
\left( \frac{1}{2b}\right) \right] -\frac{1}{2b^{2}}\right\} \,,  \notag \\
Q^{\left( 2\right) } &=&q\left( \frac{\alpha }{3\pi }\right)
\int_{0}^{\infty }d\tau \frac{e^{-\tau /b}}{\tau }\left\{ \left( \frac{1}{%
\tau }-\frac{\tau }{3}-\frac{2\tau ^{3}}{15}\right) \coth \tau -\left( 1+%
\frac{\tau ^{2}}{3}\right) \frac{1}{\sinh ^{2}\tau }\right\}  \notag \\
&=&q\left( \frac{\alpha }{3\pi }\right) \left\{ \frac{4b^{3}}{15}+\frac{b}{3}%
+\frac{2}{3}\psi \left( \frac{1}{2b}\right) -8\zeta ^{\prime }\left( -1,%
\frac{1}{2b}\right) -\frac{1}{15}\zeta \left( 3,\frac{1}{2b}\right) \right. 
\notag \\
&+&\left. \frac{1}{b}\left[ \log \left( \frac{b}{\pi }\right) +\frac{1}{6}%
\psi ^{(1)}\left( \frac{1}{2b}\right) +2\log \Gamma \left( \frac{1}{2b}%
\right) \right] +\frac{1}{2b^{2}}\right\} \,.  \label{apb7}
\end{eqnarray}

The representations above are useful to study the asymptotic regime for
large $b$. For example, using the expansions%
\begin{eqnarray}
\log \Gamma \left( \frac{1}{2b}\right) &=&\log b-\frac{\gamma }{2b}+\log
2+O\left( \left( \frac{1}{2b}\right) ^{2}\right) \,,  \notag \\
\psi \left( \frac{1}{2b}\right) &=&-2b-\gamma +\frac{\pi ^{2}}{12b}+O\left(
\left( \frac{1}{2b}\right) ^{2}\right) \,,  \notag \\
\psi ^{\left( 1\right) }\left( \frac{1}{2b}\right) &=&4b^{2}+\frac{\pi ^{2}}{%
6}+\frac{\psi ^{\left( 2\right) }\left( 1\right) }{2b}+O\left( \left( \frac{1%
}{2b}\right) ^{2}\right) \,,  \notag \\
\zeta ^{\prime }\left( -1,\frac{1}{2b}\right) &=&\zeta ^{\prime }\left(
-1\right) -\frac{1}{4b}\log 2\pi -\frac{1}{4b}\left( 1-\frac{1}{2b}\right)
+\int_{0}^{1/2b}dx\log \Gamma \left( x\right)  \notag \\
&=&\zeta ^{\prime }\left( -1\right) -\frac{1}{4b}\log 2\pi +\frac{1}{4b}+%
\frac{1}{2b}\log 2b+O\left( \left( \frac{1}{2b}\right) ^{2}\right) \,,
\label{apb8}
\end{eqnarray}%
the asymptotic behaviour large $b$ limit of Eqs. (\ref{apb1}) - (\ref{apb4})
read%
\begin{eqnarray}
&&\mathfrak{L}_{\mathfrak{F}}^{\left( 0\right) }\sim \frac{\alpha }{2\pi }%
\left( \frac{2}{3}\log b+\mathcal{K}_{\mathfrak{F}}^{\left( 0\right)
}\right) \,,\ \ \mathfrak{L}_{\mathfrak{F}}^{\left( 2\right) }\sim \frac{%
\alpha }{2\pi }\left( -\frac{b}{3}+\mathcal{K}_{\mathfrak{F}}^{\left(
2\right) }\right) \,,  \notag \\
&&\mathfrak{L}_{\mathfrak{G}}^{\left( 1\right) }\sim -\kappa \frac{\alpha }{%
2\pi }\left( -\frac{2}{3}b+\mathcal{K}_{\mathfrak{G}}^{\left( 1\right)
}\right) \,,\ \ \left( \overline{\mathfrak{F}}+\sqrt{\overline{\mathfrak{F}}%
^{2}+\overline{\mathfrak{G}}^{2}}\right) \mathfrak{L}_{\mathfrak{FF}%
}^{\left( 0\right) }\sim \frac{\alpha }{3\pi }\,,  \notag \\
&&\left( \overline{\mathfrak{F}}+\sqrt{\overline{\mathfrak{F}}^{2}+\overline{%
\mathfrak{G}}^{2}}\right) \mathfrak{L}_{\mathfrak{FF}}^{\left( 2\right)
}\sim \frac{\alpha }{2\pi }\left( b+\mathcal{K}_{\mathfrak{FF}}^{\left(
2\right) }\right) \,,  \notag \\
&&\left( \overline{\mathfrak{F}}+\sqrt{\overline{\mathfrak{F}}^{2}+\overline{%
\mathfrak{G}}^{2}}\right) \mathfrak{L}_{\mathfrak{GG}}^{\left( 0\right)
}\sim \frac{\alpha }{2\pi }\left( \frac{2b}{3}+\mathcal{K}_{\mathfrak{GG}%
}^{\left( 0\right) }\right) \,,\ \ \left( \overline{\mathfrak{G}}\mathfrak{L}%
_{\mathfrak{FG}}\right) ^{\left( 2\right) }\sim \frac{\alpha }{2\pi }\left( -%
\frac{2}{3}b+\mathcal{K}_{\mathfrak{FG}}^{\left( 2\right) }\right) \,, 
\notag \\
&&\left( \overline{\mathfrak{F}}+\sqrt{\overline{\mathfrak{F}}^{2}+\overline{%
\mathfrak{G}}^{2}}\right) \mathfrak{L}_{\mathfrak{GG}}^{\left( 2\right)
}\sim \frac{\alpha }{2\pi }\left( \frac{8b^{3}}{15}-\frac{5b}{3}+\mathcal{K}%
_{\mathfrak{GG}}^{\left( 2\right) }\right) \,,  \label{apb9}
\end{eqnarray}%
as $b\rightarrow \infty $. The $\mathcal{K}$'s are constants%
\begin{eqnarray*}
&&\mathcal{K}_{\mathfrak{F}}^{\left( 0\right) }=-\frac{1}{3}+\frac{2}{3}\log
2+8\zeta ^{\prime }\left( -1\right) \,,\ \ \mathcal{K}_{\mathfrak{F}%
}^{\left( 2\right) }=-\frac{2}{3}\gamma -8\zeta ^{\prime }\left( -1\right)
\,, \\
&&\mathcal{K}_{\mathfrak{G}}^{\left( 1\right) }=\frac{1}{3}\left( 1-2\gamma
\right) -8\zeta ^{\prime }\left( -1\right) \,,\ \ \mathcal{K}_{\mathfrak{FF}%
}^{\left( 2\right) }=-\frac{2}{3}+\frac{8}{3}\gamma +32\zeta ^{\prime
}\left( -1\right) \,, \\
&&\mathcal{K}_{\mathfrak{GG}}^{\left( 0\right) }=-\frac{1}{3}+\frac{2}{3}%
\gamma +8\zeta ^{\prime }\left( -1\right) \,,\ \ \mathcal{K}_{\mathfrak{GG}%
}^{\left( 2\right) }=\frac{2}{3}-40\zeta ^{\prime }\left( -1\right) -\frac{4%
}{15}\psi ^{\left( 2\right) }\left( 1\right) -\frac{10}{3}\gamma \,, \\
&&\mathcal{K}_{\mathfrak{FG}}^{\left( 2\right) }=\frac{2}{3}-\frac{4}{3}%
\gamma -16\zeta ^{\prime }\left( -1\right) \,,
\end{eqnarray*}%
and $\gamma \simeq 0.577$ is the Euler constant.

\end{appendices}

\end{document}